\documentclass[12pt]{article}
\usepackage{graphicx}
\usepackage[colorlinks=true, allcolors=blue]{hyperref}
\usepackage{lmodern}
\usepackage{amsmath}
\usepackage{amsfonts}
\usepackage{amssymb}
\usepackage{bm}
\usepackage{txfonts}
\usepackage[T1]{fontenc}
\usepackage[utf8]{inputenc} 
\usepackage[american]{babel} 
\usepackage{enumitem}
\usepackage[noadjust]{cite} 
\usepackage{etoolbox}
\usepackage{mathtools}

\usepackage{accents}

\graphicspath{ {figures/} }

\usepackage{amsbsy}
\usepackage{subcaption}
\usepackage{xcolor}
\usepackage{indentfirst}
\usepackage{lineno}
\usepackage{cite}
\usepackage{titletoc}
\usepackage{bigints}

\usepackage{relsize}
\usepackage{microtype}
\usepackage{array}
\usepackage{listings}
\usepackage{textcomp}
\usepackage{booktabs}

\usepackage{geometry} 
\usepackage{fancyhdr} 
\usepackage{lastpage} 
\usepackage{titlesec}
\titleformat*{\section}{\large\bfseries\upshape\sffamily}
\titleformat*{\subsection}{\large\bfseries\sffamily}
\titleformat*{\subsubsection}{\large\bfseries\sffamily}

\usepackage{times}

\usepackage{caption} 
\DeclareCaptionLabelSeparator{colon}{: }

\makeatletter
\renewcommand{\fnum@figure}{FIGURE~\thefigure}
\renewcommand{\fnum@table}{TABLE~\thetable}
\makeatother

\patchcmd\thebibliography
 {\labelsep}
 {\labelsep\itemsep=0pt\parsep=0pt\relax}
 {}
 {\typeout{Couldn't patch the command}}

\usepackage[nodisplayskipstretch]{setspace}

\makeatletter
\newcommand{\removeifnextchar}[3]{%
  \begingroup
  \ltx@LocToksA{\endgroup#2}%
  \ltx@LocToksB{\endgroup#3}%
  \ltx@ifnextchar{#1}{%
    \def\next{\the\ltx@LocToksA}%
    \afterassignment\next
    \let\scratch= %
  }{%
    \the\ltx@LocToksB
  }%
}
\makeatother
\newcommand*{\removePeriod}{%
  \removeifnextchar{.}{}{}%
}

\newcommand*{\affmark}[1][*]{\textsuperscript{#1}}

\newcommand{\doi}[1]{{doi: \href{https://doi.org/#1}{#1}}\removePeriod}

\makeatletter
\def\blfootnote{\gdef\@thefnmark{}\@footnotetext}
\makeatother

\newcommand{\xauthor}[1]{\textbf{\textsf{#1}}}

\usepackage{comment}

\usepackage{soul, xcolor}
\usepackage{array,multirow}
\usepackage{booktabs}
\usepackage[e]{esvect}

\usepackage[export]{adjustbox}
\newcolumntype{P}[1]{>{\centering\arraybackslash}p{#1}}

\definecolor{lightgray}{gray}{ .75}
\usepackage[nodisplayskipstretch]{setspace} 
\usepackage{diagbox} 
\allowdisplaybreaks

\makeatletter
\def\blfootnote{\gdef\@thefnmark{}\@footnotetext}
\makeatother

\begin{document}

\begin{center}
 \fbox{
 \parbox{0.65\linewidth}{\centering{\textcolor{gray}{Published online in the ASME \textit{Journal of Mechanical Design}} \\ DOI - \href{https://asmedigitalcollection.asme.org/mechanicaldesign/article-abstract/doi/10.1115/1.4062978/1164469/Enumeration-and-Identification-of-Unique-3D?redirectedFrom=fulltext}{https://doi.org/10.1115/1.4062978}}}} 
 \end{center}

 \begin{center}
 \large
 \textsf{\textbf{Enumeration and Identification of Unique 3D  Spatial Topologies of Interconnected Engineering Systems Using Spatial Graphs}}
 \end{center}

 \begin{center}\normalsize
 \xauthor{Satya R. T. Peddada}\affmark[1a], \xauthor{Nathan M. Dunfield}\affmark[2a], \xauthor{Lawrence E. Zeidner}\affmark[b],  \xauthor{Zane R. Givans}\affmark[1a], \xauthor{Kai A. James}\affmark[3c], \xauthor{James T. Allison}\affmark[1a] \\
 \textsf{\affmark[1]Department of Industrial and Enterprise Systems Engineering}\\
  \textsf{\affmark[2]Department of Mathematics}\\
 \textsf{\affmark[3]Daniel Guggenheim School of Aerospace Engineering}\\
 \textsf{\affmark[a]University of Illinois at Urbana-Champaign, Urbana, IL 61801}\\
 \textsf{\affmark[b]Raytheon Technologies Research Center, East Hartford, CT 06108}\\
 \textsf{\affmark[c]Georgia Institute of Technology, Atlanta, GA 30332}\\
 \textsf{Email:} \texttt{\{\href{mailto:speddad2@illinois.edu}{speddad2},\href{mailto:nmd@illinois.edu}{nmd},\href{mailto:zgivans2@illinois.edu}{zgivans2},\href{mailto:jtalliso@illinois.edu}{jtalliso}\}@illinois.edu}; \texttt{\href{mailto:lawrence.zeidner@rtx.com}{lawrence.zeidner}@rtx.com};
 \texttt{\href{mailto:kai.james@gatech.edu}{kai.james}@gatech.edu}
 \end{center}

\begin{abstract} 
{\it Systematic enumeration and identification of unique 3D spatial topologies of complex engineering systems (such as automotive cooling systems, electric power trains, satellites, and aero-engines) are essential to navigation of these expansive design spaces with the goal of identifying new spatial configurations that can satisfy challenging system requirements. However, efficient navigation through discrete 3D spatial topology (ST) options is a very challenging problem due to its combinatorial nature and can quickly exceed human cognitive abilities at even moderate complexity levels. This article presents a new, efficient, and scalable design framework that leverages mathematical spatial graph theory to represent, enumerate, and identify distinctive 3D topological classes for a generic 3D engineering system, given its system architecture (SA) --- its components and their interconnections. First, spatial graph diagrams (SGDs) are generated for a given SA from zero to a specified maximum number of interconnect crossings. Then, corresponding Yamada polynomials for all the planar SGDs are generated. SGDs are categorized into topological classes, each of which shares a unique Yamada polynomial. Finally, within each topological class, 3D geometric models are generated using the spatial graph diagrams (SGDs) having different numbers of interconnect crossings. Selected case studies are presented to illustrate the different features of our proposed framework,  including an industrial engineering design application: ST enumeration of a 3D automotive fuel cell cooling system (AFCS). Design guidelines are also provided for practicing engineers to aid the application of this framework to different types of real-world problems such as configuration design and spatial packaging optimization.
\blfootnote{Part of this work~\cite{Peddada2021} has been published previously in the proceedings of the ASME 2021 International Design Engineering Technical Conferences.}
}
\end{abstract}

\section{INTRODUCTION}
 Engineering systems design~\cite{Goode1957, Sydenham2003, Field2006} often involves choosing the most suitable candidate among many alternative design solutions to meet specific system performance criteria and design constraints using techniques such as comparative design analysis and optimization~\cite{Kim1991, Ashrafiuon1990, Zhang2011, Sergiy2018}, and configuration design~\cite{Liu2019, Blouin2004, Snavely1993, Jiang2021, Schmidt1998, Deng2022, Kott1992, Campbell2000, Grignon2004, Sigurdarson2022} methods. In most engineering design efforts, the component technologies and the component-to-component connectivity (referred to here as \textit{system architecture}) remains fixed to preserve the functionality of the system while different feasible system spatial layouts are explored. More precisely, \textit{System Architecture (SA)} specifies what components comprise a system, and how ports on components are connected to specific ports on other components. Distinct SAs represent specific technology options to perform each desired function, and define the flow paths of material, energy, and/or information from one component to another. System architecture enumeration problems have been studied previously for engineering design examples such as hybrid electric power trains~\cite{Jiang2021, Bayrak2016, Ai2005, Ramdan2016, Deng2022}, automotive vehicle suspension systems~\cite{Herber2019}, design of structures~\cite{Shim1998}, plate heat exchangers~\cite{Gut2004}, planar robot configurations~\cite{Martins2009}, and optimal cooling system layouts~\cite{Peddada2019a, Peddada2018Con} for dynamic thermal management. However, for each system architecture, many \textit {spatial topologies (STs)} may exist, each with its own optimal geometry.\\

\begin{figure}
\centering
	\begin{subfigure}{0.40\linewidth}
		\includegraphics[width=1.0\linewidth]{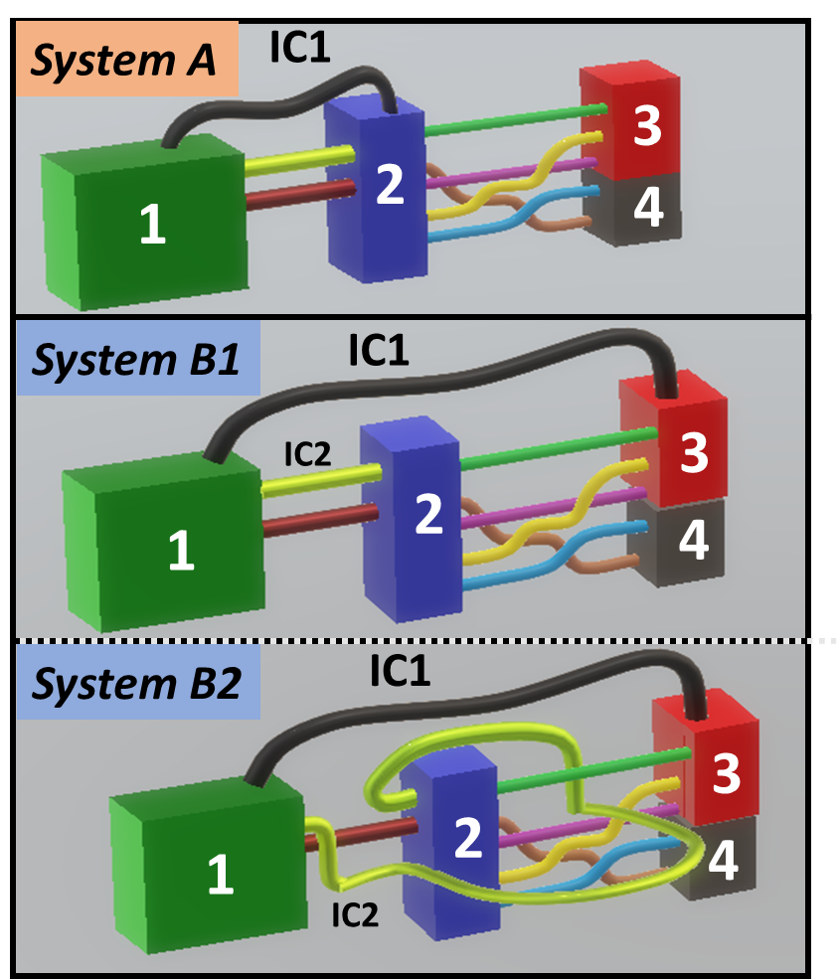}
		\caption{}
		\label{Figure1a}
	\end{subfigure}
	\begin{subfigure}{0.40\linewidth}
\includegraphics[width=1.0\linewidth]{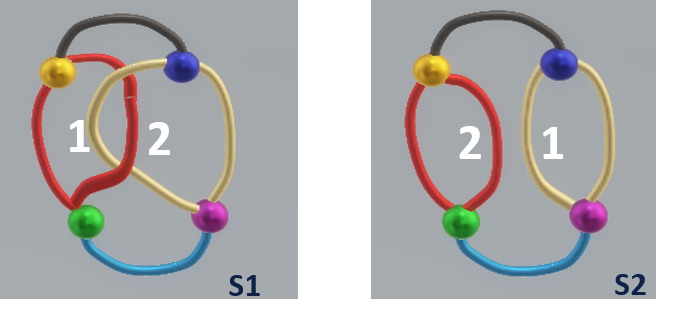}
		\caption{}
		\label{Figure1b}
	\end{subfigure}
	\caption{Illustration of system architecture and spatial topology concepts. Figure 1(a) shows moderately complex 3D systems A and B1 with different system architectures. Systems B1 and B2 have the same system architecture but are two distinct spatial topologies. Similarly, in Fig.~1(b), S1 and S2 are two different spatial topologies having the same 3D system architecture (as component-component connectivity remains the same). Observe that in S1, interconnects 1 and 2 are entangled (linked) together, whereas in S2 they are free.} .
	\label{SysArch}
\end{figure}

Any 3D engineering system geometric optimization problem can be simplified by decomposing the problem into two stages: 1) identification of unique spatial topologies, and then 2) geometrically optimizing each system configuration, accounting for physical interactions and subject to all relevant geometric, failure mode, and other constraints. For example, if an interconnect links ports $P1$ and $P2$, many options exist for how this interconnect passes around various other interconnects and components in the system. Two spatial topologies are \textit{equivalent} when there is a continuous deformation of component locations and interconnect trajectories that takes one topology to the other. This continuous deformation must be possible without cutting and rejoining any interconnects. Figure~\ref{Figure1a} shows 3D systems A and B1  having two different system architectures because interconnect IC1 is connected between components $\{1, 2\}$ in A but between $\{1, 3\}$ in B1. In other words, component-to-component connectivity is different in A and B1, respectively.  Systems B1 and B2 have the same system architecture as all the component-to-component interconnections are the same. However, B1 and B2 have different spatial topologies because the interconnect IC2 between the component 1 and 2 is topologically different (please see the crossing patterns in Fig.~\ref{Figure1a}). As both the ends of the interconnect IC2 are fixed, it cannot be continuously morphed between B1 and B2. Hence, B1 and B2 are topologically different systems. Similarly, Fig.~\ref{Figure1b} is another example. The scope of this article encompasses the challenge of enumerating and identifying such unique spatial topologies for each system architecture within a design problem. Other recent articles have focused on continuous optimization of systems once spatial topology is specified (see Refs.~\cite{Peddada2020_JMD_2Stage, Bhattacharyya2022_AIAA, Jessee2020a, Peddada2022a}, for example). The example shown in Fig.~\ref{SysArch} is kept simple for illustration purposes, but the framework can be used to generate STs for more complex architectures and larger interconnected systems with multiple crossings.\\ 

\subsection{Need for Efficient Spatial Topology Exploration Methods: To Address Fundamental Engineering Design Research Questions}\label{RQ}
 
  System architecture (SA) design automation methods studied in Refs.~\cite{Bayrak2016, Bayrak2016b, Herber2019, Peddada2019a, Muenzer2017a} have largely ignored spatial topology decisions. For instance, Refs.~\cite{Bayrak2016, Bayrak2016b} focused on simultaneous configuration and sizing optimization of automotive powertrains but not on the spatial topological invariance of the connections between the components. From an engineering design point of view, there is only a little or practically no emphasis on leveraging the "spatial topological" aspects of these 3D systems during the design exploration or optimization process. How are the components and interconnects oriented or interacting among themselves \textit{in real 3D space}? How are they located in that 3D space relative to each other? For example, do two interconnects cross each other and how are their physical fields affecting components or other interconnects in their proximity? What engineering design representations would be suitable to capture the spatial  interactions (or the spatial topology) between the different system elements? There are several fundamental research questions that need to be carefully and thoroughly addressed. In other words, studying spatial dimensions and spatial interactions is very important as it directly has practical implications on the amount of 3D volume occupied by the engineering system, the accessibility of its components and how each of the subsystems in turn get influenced by their multi-physics spatial fields such as heat losses, electromagnetic radiation, and thermal stress or mechanical fatigue acting within the system.\\
  
  For example, there exists a lot of wiring/piping, and components such as a battery, electric  motor, power converter, etc. in a hybrid electric vehicle (HEV) system. The spatial locations and topologies of these components and interconnects contribute to the overall spatial packaging density (SPD) of the HEV. SPD directly influences the vehicle's efficiency, road range, and total energy consumption. Two HEV system models might have the same system architecture and component connectivity, however, their performance will be completely different because of the spatial locations and arrangements of its components and interconnect network. If lesser volume is occupied, heat loss radiation, noise and vibration, and cooling requirements should be addressed more aggressively. Furthermore, compactness makes it challenging to address accessibility and maintenance issues. A maintenance engineer would be glad if there is lot of working space available between components and subsystems with lesser entangled wiring for performing efficient overhaul, repair, or replacement operations. But the customer and product designer may want a compact vehicle owing to better efficiency, road range, and energy consumption. Therefore, there are key design trade-offs that exist between different vehicle performance attributes such as overall system efficiency, packaging density, and accessibility due to spatial topological decisions. Similarly, 
 for designing automotive climate and thermal HVAC systems, sometimes the hot and cold refrigerant pipelines must not be close to each other in terms of physical proximity to avoid heat radiation effects. In such cases, spatial topological exploration is very helpful. Would it be possible to find other configurations where the pipelines are far apart while the system connectivity remains the same? It might be possible to take such decisions for one or two components or pipes. But for a large vehicle with hundreds of system elements, it is a humanly impossible task to come up with an optimal arrangement that takes into consideration different design choices. Hence, from an engineering design perspective spatial topological decisions are vital and there is a need to efficiently navigate through the discrete 3D design space to achieve optimal system configurations that balance between spatial packaging density, physics performance, and design for accessibility and maintenance.\\ 

As described above, there are many pain points in real industry practice due to 3D spatial constraints and such design challenges must be thoroughly addressed to develop efficient, sustainable, and easily maintainable systems. This article aims to arouse in-depth discussion among engineering design researchers and help the larger design community realize the need for new representations and design automation frameworks to address such complex design problems and make relatively faster and better decisions. In summary, this article is trying to address some of the following fundamental research questions as listed below:

\begin{enumerate}
    \item Does there exist a design representation that can capture 3D spatial topology of engineering systems? There has been a lot of work done in system architecture exploration. However, we need a design representation that can capture the 'spatial' relationships between components and interconnections within the 3D system. 
    \item For systems having components immersed in multi-physics spatial fields there might be different operating conditions that need to be met. For example, hot and cold fluid lines may need to stay apart, and components may be separated by bays or walls to avoid electromagnetic radiation or noise/vibration issues.
    \item Many real-world engineering systems contain hollow objects. Design decisions might include whether a pipe or duct should pass through or outside the hole.
    \item Can new algorithms be created that can navigate 3D discrete space efficiently to explore possible design candidates that satisfy operational and maintenance constraints?
    \item For a given system, how many spatial topological solutions exits for multi pipe-routing or wiring
    \item Are there mathematical ways to map abstract designs into polynomials for faster evaluation and comparison to filter non-unique spatial topologies? 
    \item What would be running time complexity for such design automation algorithms or frameworks?
    \item Would such representations support simultaneous geometry, physics, and topology optimization? Can gradient-based optimization methods be used? 
    \item Would such a spatial topology enumeration framework help maintenance engineers determine what is the optimal trajectory that components can take to be easily removed or replaced from the system. What is the system configuration that has maximum access ports for critical components? 
    \item What is best spatial configuration of piping or wiring to support tight packaging constraints?
    \item Can these generic algorithms be applied to systems at various scales?
    \item Can artificial intelligence and machine learning techniques be employed to make these methods efficient and scalable?  
\end{enumerate}

The above questions are generic and fundamental and applicable to any interconnected 3D engineering system. Some common design engineering questions have been outlined and this is not considered to be an exhaustive list. However, the decisions involved in each question can have great impact on overall system performance, early design stage process, and several life-cycle consideration. Interesting case studies have been demonstrated in Sec.~\ref{CaseStudy} as an attempt to answer some of these above research questions using the proposed enumeration framework. The inferences obtained from the detailed case studies have been presented in the discussion section~\ref{Discussion}. However, it mus be noted that this article is a preliminary attempt to answer these challenging questions and answering all questions is definitely out of scope of this article. Further investigation is being done on these topics and a lot of support is required from the larger design engineering community to make impact.

\subsection{Design Process and the Power of Design Representations}
Any systematic design process~\cite{Field1997,Challender2000} involves four key tasks: \textit{representation}, \textit{generation}, \textit{evaluation}, and \textit{design guidance}. \textit{Representation} refers to the task of describing a system using a generic model that captures the functionality of the various system elements. Depending on the design analysis tools and application requirements, design representations can be mathematical, graphical, physics-based, or conceptual~\cite{Wyatt2013a, Schmidt1999}. A formal representation of engineering systems is necessary for generation of their different spatial configurations. Previous works utilized different representations such as stick diagrams~\cite{Oraon2018}, canonical graphs~\cite{Babai1983CanonicalLO, Rensink2004}, lever diagrams~\cite{Ross1991, Liao2017},  design structure matrices (DSMs)~\cite{AlGeddawy2015}, and bond graphs~\cite{Bayrak2016, Pease1988, Wu2008, Beaman1988, Behbahani2013, Bachrach1996}. Each of these representations has advantages and disadvantages that impact the computational effort required to generate the feasible design space of engineering system configurations. The \textit{generation} phase involves creating feasible design alternatives using the representation based on design rules. Configuration synthesis and automatic enumeration strategies~\cite{Deng2022, Bayrak2016} that meet design constraints have been developed previously by researchers for applications such as planetary gear sets (PSGs) for automotive hybrid transmission systems (HTS)~\cite{Xu2019}, split hybrid configurations for single planetary gears~\cite{Barhoumi2017}, robotic manipulator configurations~\cite{Lipkin1991}, and vehicle suspension mechanisms~\cite{Liu1993}. \textit{Evaluation} is the process of measuring the design quality in terms of the performance criteria. Finally, \textit{design guidance} is providing feedback for the generation task based on the evaluation output to find better  alternatives in the design space. In design automation methods, the generation, evaluation, and design guidance tasks are performed in an automated loop that, when successful, converges to a design solution. However, the element of this process that actually enables high design accuracy, comprehensive search space navigation, and computational efficiency is the design representation that is selected before the start of this iterative process. Hence, a design representation for 3D interconnected systems that captures the relevant problem attributes and aligns well with a generation computation method is critical for navigating through the discrete 3D spatial topology options efficiently.\\ 

The design elements of any three-dimensional interconnected engineering system are its components, their 3D spatial locations and orientations, their port valencies, interconnections (diameters/sizes and trajectories), and the crossings between their interconnections. Unlike 2D system layout enumeration and optimization performed in the design of very large scale integrated circuits (VLSI)~\cite{Sharma2014_VLSI_OptimizationSurvey, Devadas1995ASO, Agnesina2020VLSIPP, BakerAlawieh2019GenerativeLI, Bunglowala2009OptimizationOH, Jung2016OpenDesignFD, Nath2015ANA, Geetha2017DesignMA, Kumar2020ReviewOV, Dewan2020NPSeparateAN, Held2011CombinatorialOI, Kaeslin2014TopDownDV}, the 3D spatial layout enumeration problem is fundamentally different and an even more challenging problem as described in Ref.~\cite{Peddada2021b, Peddada2022a, peddada2023automated}. More specifically, even for a given system architecture, options exist for how the interconnects are routed relative to one another and to the components (for example, say, should duct A go over cable B or under, or should a pipe be routed through the hole in the casing, or around the edge of the casing, etc.). These are topologically discrete design differences. To cater to that need, in this article we have identified a viable mathematical design representation to describe 3D interconnected spatial layouts as \textit{spatial graph diagrams (SGDs)}. 

\subsection{Objectives and Contributions}
Complex engineering systems such as autonomous aerial vehicles~\cite{Moguel2017}, electric power trains~\cite{Bayrak2016, Yu2018}, aero-jet engines~\cite{Kyprianidis2017AnAT}, or vehicle thermal management and cooling systems~\cite{Park2010, Peddada2019a} have different kinds of components connected together either through wires, ducts, or pipes entangled with one another in a tightly packed three-dimensional volume. Engineering design alternatives with distinct spatial topologies may exist, with different values of metrics such as efficiency, spatial packaging density, maintenance costs, and design complexity. Current practice for exploring different system spatial topologies relies largely upon  human  expertise, design rules, creating new designs derived from existing ones (resulting in incremental design advancements), and manual adjustments (limiting complexity of designable systems). 

This approach bounds the pace and scale at which spatial topologies can be explored for practical application to typical complex systems. The realization of significantly enhanced functionality or performance is prevented by the profoundly incomplete design space coverage of current practice. More efficient automated methods are needed to unleash engineering system design efforts from the restraints of current design practice for spatial topological decisions. Enumeration methods in low-dimensional topology for knots are well known, which are a special class of spatial graphs, see \cite{HosteThistleWeeks1998} for an overview.  Using additional techniques from hyperbolic geometry, it is possible to exactly enumerate all knot topologies with less than 20 crossings, of which there are more than 350 million \cite{Burton2020}. Compared to such massive computations, prior work on spatial graphs where the underlying graph is more complicated than just a loop is limited: mostly tabulations of less than 100 topologies \cite{Oyamaguchi2015EnumerationOS, Kanenobu2012FiniteTI, Moriuchi2008EnumerationOA, Moriuchi2009ATO, Soma1996SpatialgraphIF, FominykhEtAl2016}. For example, the authors in Ref.~\cite{Moriuchi2009ATO} generated two vertex bouquet spatial graphs with a maximum of seven crossings. We have taken inspiration from works in mathematical low-dimensional topology to create new algorthms that can generate 3D system spatial topologies using spatial graphs. As our cases studies in Sec.~\ref{CaseStudy} demonstrate, the enumeration strategy in this article allows for much larger-scale enumerations, with arbitrary specified system architectures.  This is very suitable for representing large-scaled complex engineering systems easily, and for enumerating their 3D spatial topologies efficiently.
\\ 

The main objectives of the work presented in this article are: 1) to present fundamental engineering design research questions listed in Sub-section~\ref{RQ}, and 2) to develop an automated and systematic enumeration framework to both represent 3D engineering systems using spatial graph diagrams (SGDs) and efficiently generate distinct spatial topologies (STs) for a given system architecture using a rigorous mathematical approach as an attempt to address the research questions outlined at the end of Sub-section.~\ref{RQ}.

The major contributions of the proposed design framework include:  
\begin{enumerate}
      \item A new way to represent 3D engineering system configurations using spatial graph theory. This representation supports description of components as nodes, interconnects as edges, allows multiple interconnect crossings, and variable component valency. 
      \item Combinatorial enumeration of all spatial graph diagrams (SGDs) for a given system architecture up a maximum number of interconnect crossings. The proposed method presented in this article supports enumeration of spatial topologies for moderately-scaled engineering system architectures containing (approximately) up to 20 components, 60 interconnects, and 10 crossings. To enumerate spatial topologies for much larger scale systems, an effective decomposition-based method, such as the approach presented in Case Study~\ref{CS5}, can be utilized. As part of future work, we also plan to investigate alternative methods such as deep learning and pattern recognition as mentioned in Ref.~\cite{Davies2021} to efficiently explore newer topologies using large-scaled system data sets of 3D system architectures. This sort of strategy has been used previously to scale SA design to problems too large for enumeration using machine learning strategies in conjunction with enumerated data as articulated in Refs.~\cite{Guo2018d, Guo2019a, Guo2018PhD, Parrott2023a}  
      \item Efficient and systematic identification of unique SGDs from the exhaustively enumerated SGD set using Yamada polynomial invariants. The Yamada polynomials are used as a tool to identify any duplicate spatial graph topologies, similar to isomorphisms for standard graphs. This serves as a foundation to explore the discrete 3D spatial topology design space thoroughly. 
      \item Topological equivalence between spatial graphs is tested here using Yamada polynomials rather than comparing graph diagrams directly. 
      \item Practical demonstration of the proposed design framework on a real-world industry application: enumeration of unique spatial topologies of an automotive fuel cell system (AFCS). 
      \item Engineering design insights and guidelines to help system design engineers apply this proposed framework to different kinds of practical applications. An in-depth discussion has been presented in Sec.~\ref{Discussion}.
\end{enumerate}

The remainder of this article is organized as follows. The terminology and notation of the proposed spatial graph representation are introduced in Sec.~\ref{SpatialGraphs}. Section~\ref{Yamada} describes the characteristics of Yamada polynomials and how they are evaluated for an individual SGD. In summary, Secs.~\ref{SpatialGraphs} and~\ref{Yamada} below provide a detailed discussion introducing readers to the mathematical theory of spatial graphs, Yamada polynomial invariants, and graph relations through a few illustrative examples.  
 Section~\ref{Framework} demonstrates the proposed six-step design framework that utilizes spatial graphs to represent, enumerate, and identify distinctive 3D topological classes for an engineering system, given its system architecture (SA). SGDs are generated for a given SA from zero to a specified maximum crossing number. Corresponding Yamada polynomials for all the enumerated SGDs are then generated. SGDs are categorized into topological classes, each of which shares a unique Yamada polynomial. Finally, for each topological class, 3D geometric models are generated that can be used for multiphysics configuration design optimization. Section~\ref{CaseStudy} presents several interesting practical case studies based on the proposed framework. The results are discussed in Sec.~\ref{Discussion} from a engineering systems design perspective. Finally, the conclusion and future work items are presented in Sec.~\ref{Conclusion}.

\section{SPATIAL GRAPHS} \label{SpatialGraphs}

\begin{figure}
		\centering
		\includegraphics[width=3.3in]{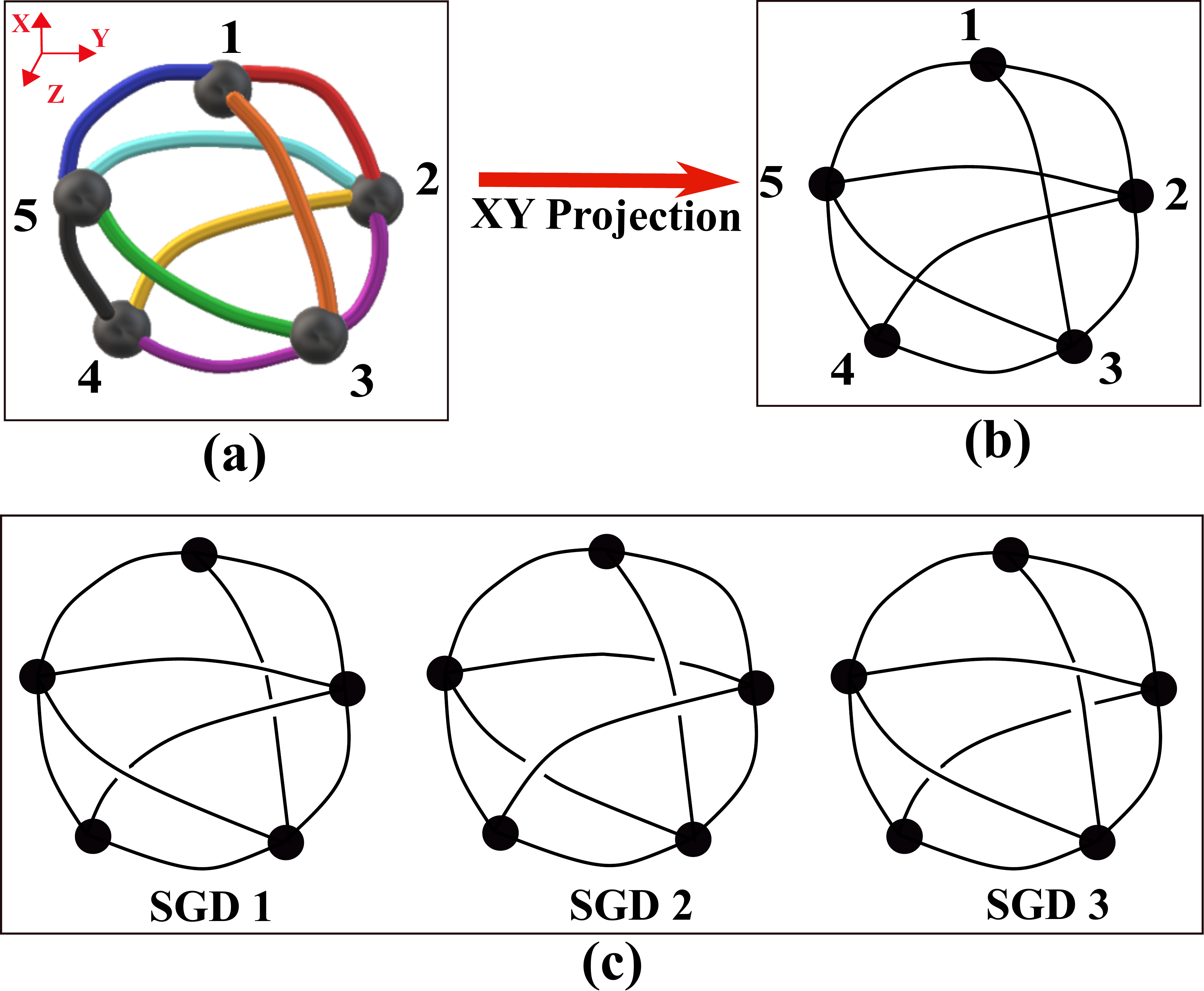}
		\caption{Figure \textbf{(a)} represents a five component interconnected 3D system; \textbf{(b)} is the shadow of the 3D diagram; and \textbf{(c)} shows three distinct spatial graph diagrams  (SGD1, SGD2, and SGD3)}
		\label{SG_Representation}
\end{figure}

The study of graphs in 3-space has been mathematically formalized using \textit{spatial graphs} \cite{Flapan2016RecentDI, Taylor2019AbstractlyPS, Flapan2012SpatialGW}, which we now describe.  Suppose $G$ is a graph, that is, a set of vertices and a set of edges, where an edge is just a pair of vertices.  (Edges are undirected and multiple edges between the same pair of vertices are allowed.)  A \textit{spatial embedding} of a graph $G$ is a set of points (nodes) in $\mathbf{R}^{3}$ corresponding to the vertices of $G$, and a set of smooth arcs (links) corresponding to the edges of $G$ that join appropriate pairs of vertices; here, each arc meets the vertices only at its two endpoints, and it intersects other arcs only at these vertices.  Collectively, these points and arcs form a \textit{spatial graph} with underlying (abstract) graph $G$.  More formally, the spatial embedding is a function $f:G \rightarrow \mathbf{R}^{3}$, whose image $\tilde{G}:= f(G)$ is the spatial graph.  See Figure~\ref{SG_Representation}(a) for a sample spatial graph.  The natural topological notion of equivalence for spatial graphs is \textit{isotopy}, when two spatial graphs $\tilde{G}_1$ and $\tilde{G}_2$ can be continuously deformed from one to the other without any arc passing through another arc or itself. \\

Spatial graphs are a natural extension of knot theory, which is the study of circles embedded in $\mathbf{R}^{3}$, since we can put vertices on a knot to make it into a spatial graph. While the study of knot theory has its origin in the physics of the late 19th century \cite{HosteThistleWeeks1998}, spatial graph theory has its roots in chemistry~\cite{Liang1994, Flapan2013} and is different from graph theory because graph theory studies abstract graphs while spatial graph theory studies embeddings of graphs in $\mathbf{R}^{3}$ or even in other 3-manifolds~\cite{FLAPAN1995, Mellor2018InvariantsOS, Fleming2006AnIT}. This theory was used in polymer stereochemistry~\cite{Liang1994, Rapenne2000}, ecology research~\cite{Dale2017}, and molecular biology (e.g.,~protein classification and protein structure prediction~\cite{Tunyasuvunakool2021, Song2022, Heal2018, Huan2003MiningSM} using deep learning) to distinguish different topological isomers. A folded protein can be thought of as a spatial graph where residues are the nodes and edges connect the residues in close proximity. This article's goal is to improve and tailor the immense potential of these representations in developing new engineering design automation methods.\\

If a spatial graph is projected onto a plane, then some arcs (edges) may appear to cross in the projection. If information about which arc is on top at the apparent crossings is omitted, the projection is called a \textit{shadow} of the spatial graph, as shown in Fig.~\ref{SG_Representation}(b). If we keep track of which arc is on top at each apparent crossing, the projection or planar representation is called a \textit{diagram} of the spatial graph, as shown in Fig.~\ref{SG_Representation}(c). In other words, diagrams are the images of embedded graphs under a projection $\mathbf{R}^{3} \rightarrow \mathbf{R}^{2}$ whose singularities are a finite number of crossings of edges equipped with over-under crossing information. Hence, many different spatial diagrams of a spatial graph $G$ may have the same shadow. We can produce this family of spatial graph diagrams (SGDs) by assigning all possible permutations of overstrand or understrand information.

\subsection{Reidemeister moves}
\begin{figure}
		\centering
		\includegraphics[width=3in, height  = 4in]{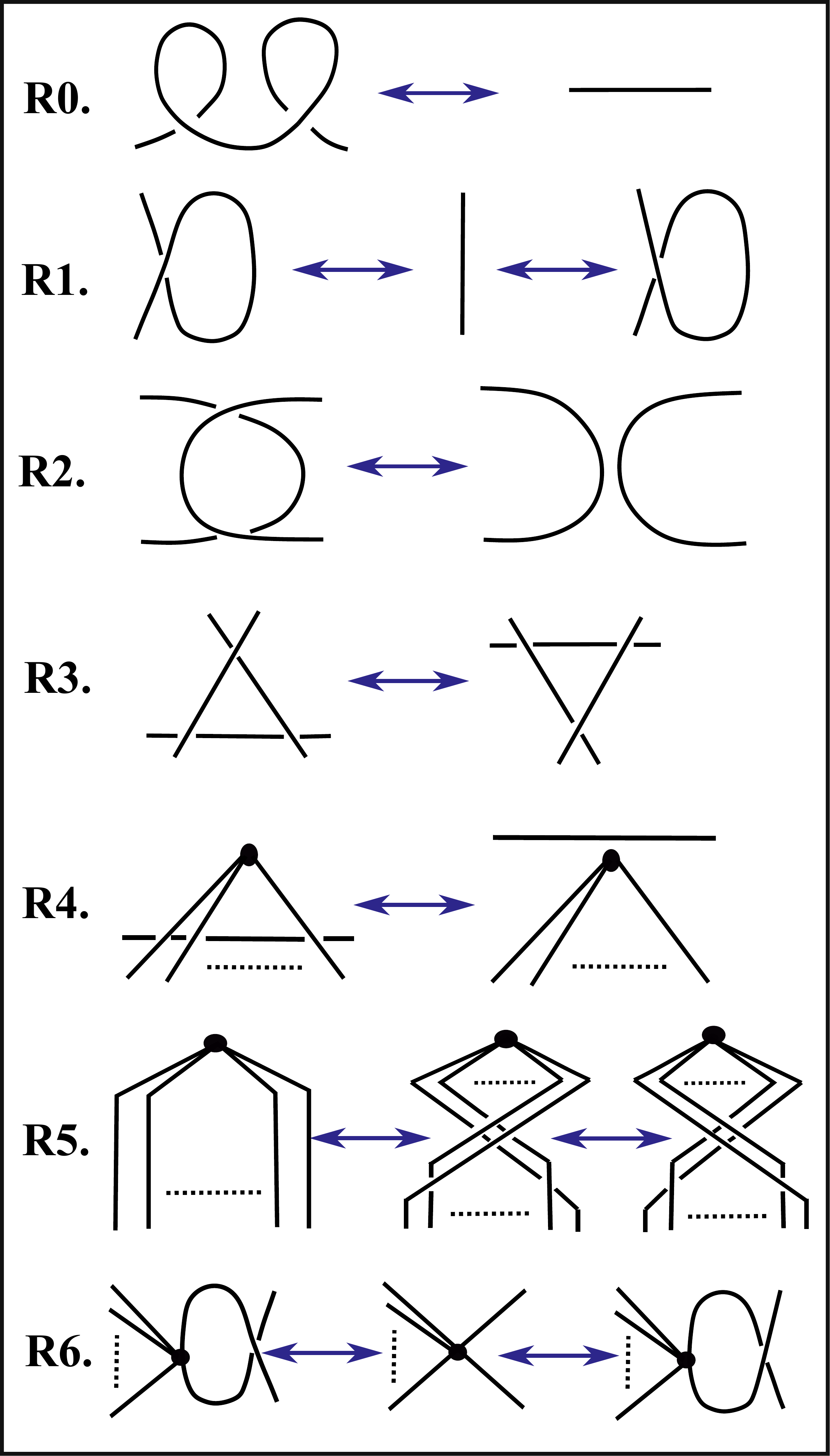}
		\caption{Fundamental Reidemeister moves for spatial graphs. R5 describes the move of taking the node and rotating along its vertical axis, dragging the strands. In R6 move, only the two strands to the right of the node are being moved.}
		\label{Reidemeister_moves}
\end{figure}

\begin{figure}
		\centering
		\includegraphics[width = 3.3 in]{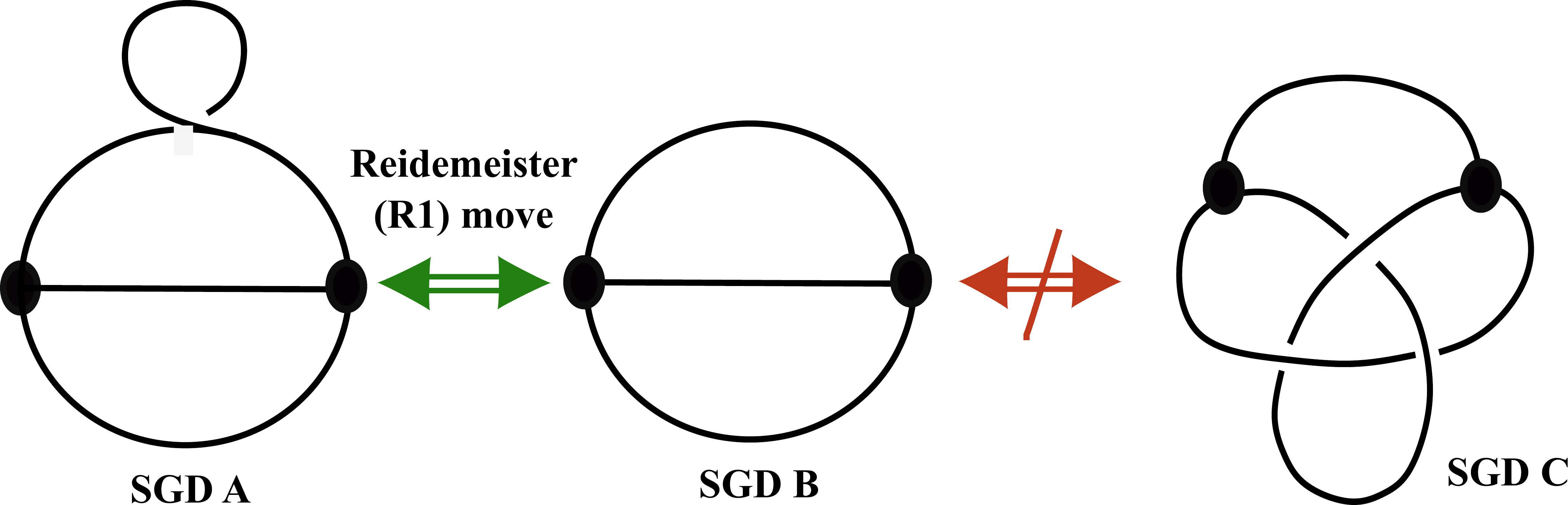}
		\caption{SGDs A and B are topologically equivalent $\theta_{1}$ graphs under Reidemeister-I (RI) move.  C is a $\theta_{2}$ graph and is topologically distinct from A and B under any fundamental R moves.
		}
		\label{R1-moveEXample}
\end{figure}

Given an abstract graph $G$, we can use the spatial graph diagrams above to begin enumerating spatial embeddings of $G$. The challenge is then to determine which of these SGDs actually describe isotopic spatial embeddings (i.e.,~are topologically equivalent), so that later steps in the design process consider each topological possibility only once.  Fortunately, it has been shown that two diagrams represent isotopic embeddings if and only if they are related by a finite sequence of fundamental \textit{Reidemeister moves ($R0$ to $R6$)}~\cite{Trace1983OnTR, Hass1998TheNO, Hayashi2005} as shown in Fig.~\ref{Reidemeister_moves}. Figure~\ref{R1-moveEXample} shows a simple illustration of three diagrams where SGDs A and B are \textit{topologically equivalent} under the first Reidemeister move $R1$ whereas C is not equivalent to either A or B as its edges cannot be continuously deformed using the Reidemeister moves to attain A or B, so they represent \textit{topologically distinct} spatial graphs.

\subsection{Flat vertex graphs and ribbon graphs} \label{FlatVertex}

\begin{figure}
		\centering
		\includegraphics[width=3.5in]{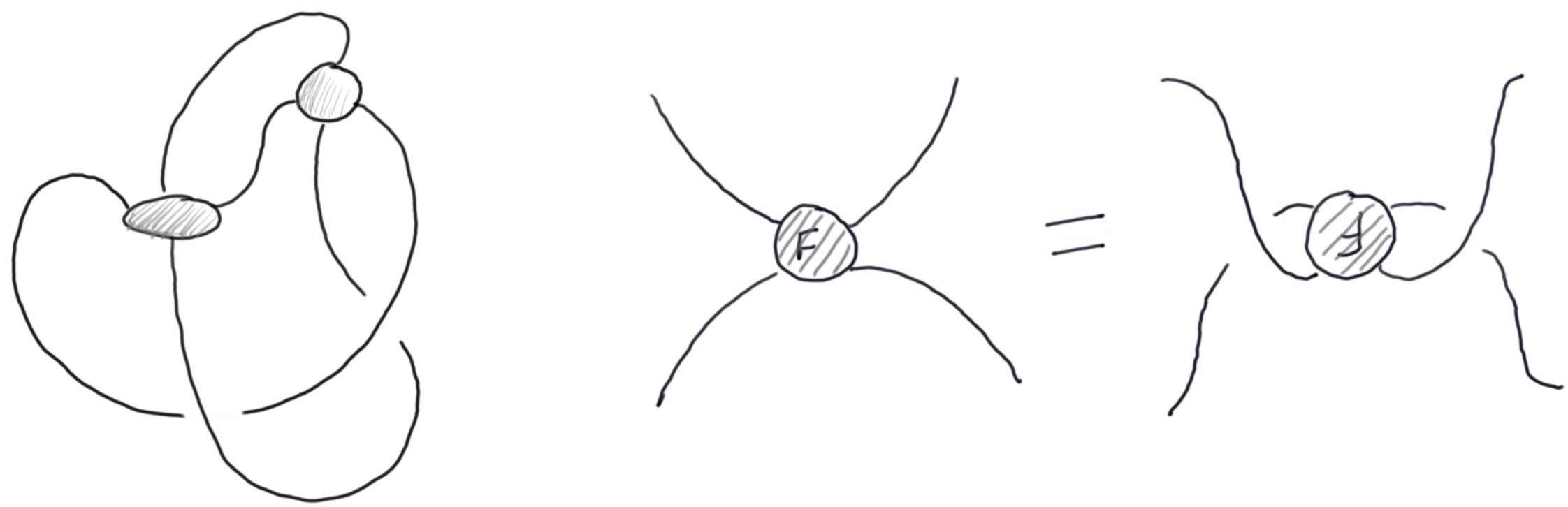}
		\caption{At left is a typical flat vertex graph.   The Reidemeister $R5$ move on this kind of spatial graph is shown at right. }
		\label{fig: flat}
\end{figure}

The topological formulation of spatial graphs is quite idealized in that each vertex has no local structure and the edges are infinitely thin.  We can impose additional, but still purely topological, structure by considering \textit{flat vertex graphs} and \textit{ribbon graphs}, which may be more suitable for certain design applications.  A flat vertex graph is a spatial graph where the vertices correspond to flat disks in $\mathbf{R}^3$ as shown in Fig.~\ref{fig: flat}.  In particular, this gives the edges coming in to each disk a cyclic order.  A flat vertex graph can also be encoded by a SGD, with the convention that each disk is rotated parallel to the projection plane before projecting.  Two SGDs represent isotopic flat vertex graphs if and only if they differ by a series of Reidemeister moves $R0$ to $R5$; here $R6$ is no longer allowed since it would change the order of the edges coming into the vertex disk.  \\

\begin{figure}
		\centering
		\includegraphics[width=3.5in]{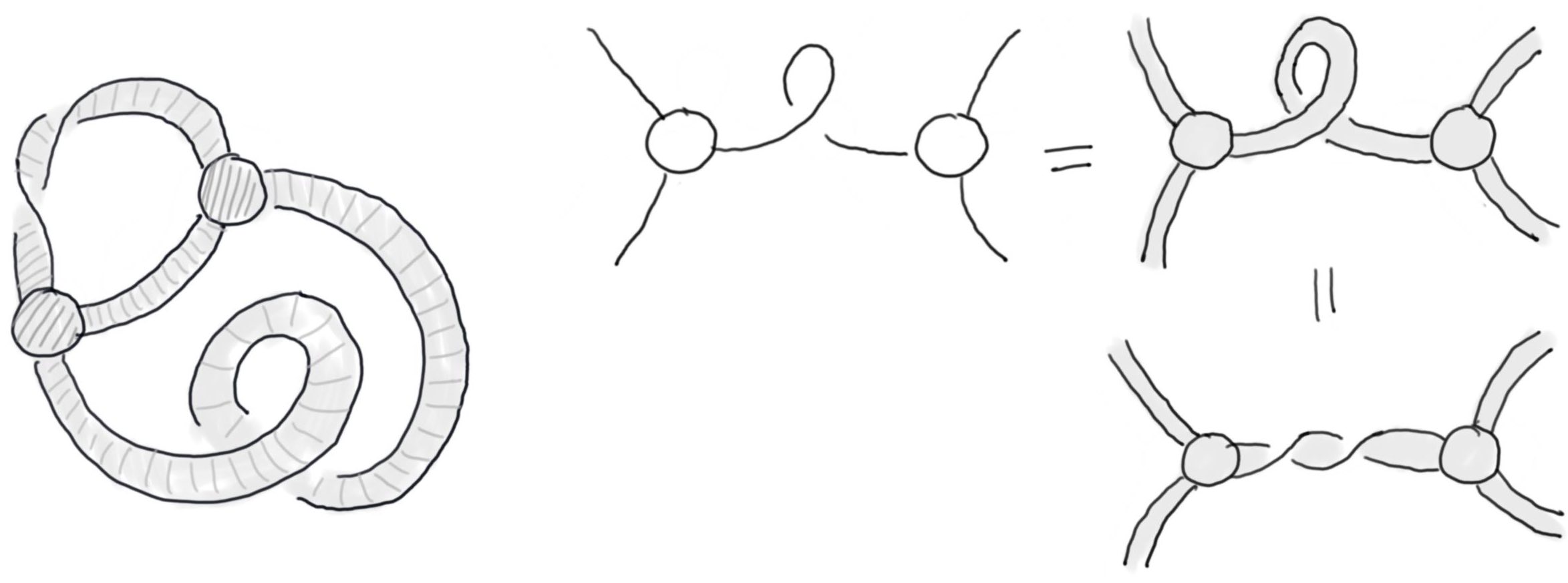}
		\caption{At left is a typical ribbon graph; notice the twists in the topmost band. The upper-right shows how a SGD encodes a ribbon graph via the blackboard framing, with the vertical isotopy showing how loops in the SGD can describe twists in the ribbons.}
		\label{fig: ribbon}
\end{figure}

A ribbon graph is a spatial graph whose vertices have become flat disks and whose edges have become thin bands, depicted in Fig.~\ref{fig: ribbon}.  These too can be encoded as SGDs by using the blackboard framing convention (a way to view a knot diagram on a plane) illustrated in Fig.~\ref{fig: ribbon}. This framing is obtained by converting each component to a ribbon lying flat on the plane.  Two SGDs represent isotopic ribbon graphs if and only if they differ by a sequence of Reidemeister moves $R0$ and $R2$ to $R5$.  The basic notation of a spatial graph introduced in Sec.~\ref{SpatialGraphs} is sometimes referred to as a \textit{pliable} spatial graph to contrast the notion with flat vertex and ribbon graphs.  Here, we will focus on pliable and flat vertex graphs, but note that ribbon graphs would be useful for measuring twisting along interconnects in the final 3D system.

\section{YAMADA POLYNOMIAL INVARIANTS} \label{Yamada}
Reidemeister moves are valuable for identifying when two embeddings are isotopic (that is, topologically equivalent); however, finding the specific sequence of moves between two equivalent spatial diagrams can be extremely challenging, especially when the spatial graphs have many nodes and edges.  Even for knots, which are the simplest class of spatial graphs, it is unknown whether there exists a polynomial-time algorithm for determining when two knots are isotopic.  (It is not impossible that such an algorithm exists: the question of whether a knot is equivalent to a round circle is in $\mathbf{NP} \cap \mathbf{coNP}$ \cite{HassLagariasPippenger1999, lackenby2019efficient}.) To show that two embeddings are not \textit{isotopic} requires an \textit{invariant}: a function of the embeddings whose output is not changed by isotopies, and which takes different values on the two embeddings~\cite{Ishii2011ONNO, Negami1987PolynomialIO, Cho2011TopologicalSG, Flapan2017KnotsLS}.  Mathematicians often use such invariants that are computable and yet powerful enough to detect some delicate differences of embeddings of the same graph. Over the last century, many \textit {polynomial invariants}~\cite{Negami1987PolynomialIO, BarNatan1995OnTV, Kauffman1988NewII, Thompson1992API} were discovered by knot theorists,  such as the Alexander-Conway~\cite{AlexanderTopologicalIO}, Jones~\cite{Murasugi1987JonesPA}, Kauffmann~\cite{Kauffman1989InvariantsOG}, and Yoshinaga~\cite{Dobrynin2003OnTY} polynomials.  Some of these have been extended to spatial graph theory~\cite{Yokota1996TopologicalIO, Kong2015ColoringsDA, Murakami1993TheYP} using similar constructions. These invariants satisfy nice \textit{skein relations} which are mathematical tools that give linear relationships between the polynomials of closely related diagrams. Relevant skein relations are sufficient to calculate the polynomials recursively and are relatively convenient to use for this purpose. The proof of invariance then relies on using the skein relation to show the value of the invariant is unchanged by Reidemeister moves.

\subsection{Yamada polynomial properties}
The specific polynomial invariant used here is the Yamada polynomial, which associates to each SGD a polynomial in an indeterminate $A$, which is an arbitrary independent variable.  For example, it turns out that the Yamada polynomial for the SGD C in Fig.~\ref{R1-moveEXample} is $-A^{-6} -A^{-5} -A^{-4}-A^{-3}-A^{-2}-1-A^{2}+A^{6}$.  The Yamada polynomial is defined in terms of a polynomial invariant $H$ of ordinary (non-spatial) graphs.  Continuing the example, the abstract theta graph that underlies all the SGDs in Fig.~\ref{R1-moveEXample} has $H(A) = -A^{-2}-A^{-1}-2-A-A^{2}$. We first discuss the polynomial invariant $H$ and its properties P1--5, which can be used to recursively compute $H$ for any graph. We then turn to the Yamada polynomial and its fundamental properties S1--4 which similarly allow for recursive computation of it for any SGD.  After giving some helpful additional properties Y1--8 of the Yamada polynomial, it is computed for four different example SGDs in Sec.~\ref{Examples}.\\

Let $G = (V,E)$ be an abstract graph, where $V$ is the vertex set and $E$ is the edge set of $G$. For two graphs $G_{1}$ and $G_{2}$, $G_{1} \sqcup  G_{2}$ denotes the disjoint union of  $G_{1}$ and $G_{2}$, and $G_{1} \vee  G_{2}$ denotes a wedge at a vertex of $G_{1}$ and $G_{2}$, that is $G_{1} \vee  G_{2} = G_{1} \cup G_{2}$  and $G_{1} \cap  G_{2} = \{\text{vertex}\}$. In addition, a graph $G$ has a cut-edge $e$ (also known as bridge or isthmus) if $G - e$ has more connected components than $G$. First, following Ref.~\cite{Yamada1989}, a polynomial invariant $H(G)(A)$ of an abstract graph $G$ is described, where $A$ is an indeterminate (arbitrary independent variable); precisely, our $H(G)(A)$ is Yamada's $h(x, y)$ with $x = -1$ and $y = -A - 2 - A^{-1}$.  The polynomial $h(G)(A)$ is  characterized by the following properties:
 \begin{enumerate}  [label={P}{{\arabic*}.}]
    \item $H(\mbox{empty graph)} = 1$ and $H(\mbox{simple loop}) = A + 1 + A^{-1}$.
    \item $H(G_{1} \sqcup G_{2}) = H(G_{1})H(G_{2})$
    \item $H(G_{1} \vee G_{2}) = - H(G_{1})H(G_{2})$
    \item If $G$ has a cut edge, then $H(G) =0$.
    \item Let $e$ be a non-loop edge of a graph $G$. Then $H(G) = H(G/e) + H(G-e)$, where $G/e$ is the graph obtained from $G$ by contracting $e$ to a point and $G-e$ is $G$ with $e$ deleted.
\end{enumerate}

 \begin{figure}
		\centering
		\includegraphics[width = 2in]{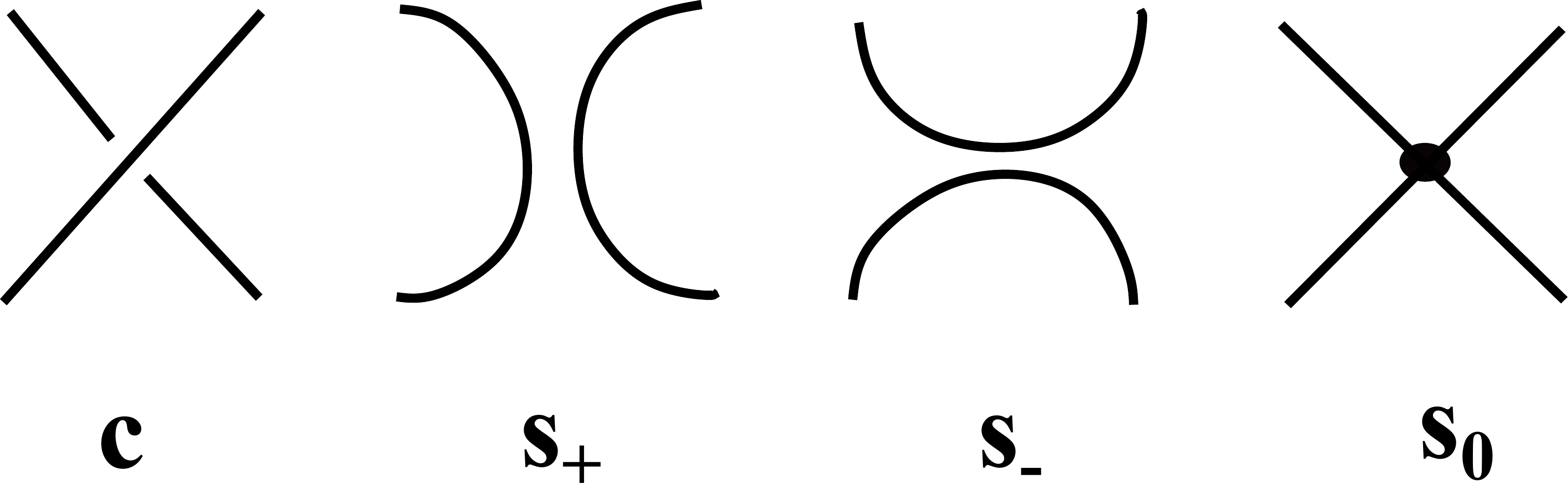}
		\caption{Class of spins for a crossing.}
		\label{Spins}
\end{figure}

 Now we can define a powerful and much-studied invariant of spatial graphs, the \textit{Yamada polynomial}~\cite{Yamada1989, Vesnin1996, Murakami1993TheYP, Li2018OnYP, Deng2018TheGY}. Let $g$ be the a spatial graph diagram. For a crossing $c$ of $g$, three \textit{graph reductions} are defined as: $s_{+}$, $s_{-}$, and $s_{0}$ (denotes a vertex), with the class of spin $+1$, $-1$, and $0$, respectively, as shown in Fig.~\ref{Spins}. These graph reductions are used to replace crossings in a spatial graph for Yamada polynomial calculation. Let $S$ be the planar graph obtained from $g$ by replacing each crossing with a spin. $S$ is called a \textit{state} on $g$ and $U(g)$ denotes the set of states on $g$ obtained by applying all possible reductions in its crossings. Set $\{g|S\} = A^{n_{1}-n_{2}}$, where $n_{1}$ and $n_{2}$ are the numbers of crossings with spin of $+1$ and spin of $-1$, respectively, and $A$ is an indeterminate. The Yamada polynomial $R\{g\}(A) \in \mathbf{Z}[A,A^{-1}]$ is defined as:
 \begin{gather*}
     R(g) = R(g)(A) = \sum_{S \in U(g) } \{g|S\}H(S),
 \end{gather*}
In particular, if the diagram of $g$ does not have crossings, then $R(g) = H(g)$.  This Yamada polynomial for a spatial graph can be computed recursively using the following skein relations and the properties of $H$:
 
\begin{enumerate} [label={S}{{\arabic*}.}]
    \item  $R({\includegraphics[width = 0.8cm, valign=c]{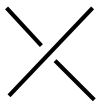}}) = AR({\includegraphics[width = 1cm, valign=c]{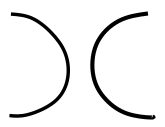}}) + A^{-1}R({\includegraphics[width = 0.8cm, valign=c]{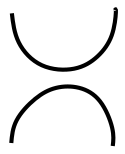}}) + R({\includegraphics[width = 0.8cm, valign=c]{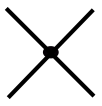}}),$
    \item $R({\includegraphics[width = 1.4cm, valign=c]{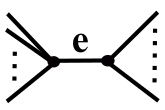}}) = R({\includegraphics[width = 1.2cm, valign=c]{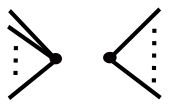}}) + R({\includegraphics[width = 1cm, valign=c]{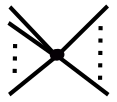}}),$ \text{where} $e$ \text{is a non-loop edge}.
    \item $R(g_{1} \sqcup g_{2}) = R(g_{1}) R(g_{2})$,
    \item  $R(g_{1} \vee g_{2}) = -R(g_{1}) R(g_{2})$.
\end{enumerate}

So far, the Yamada polynomial is a function of the given diagram $g$ and we need an invariant of the  spatial graph $\tilde G$ it describes.  Yamada showed:
\begin{enumerate}[label={I}{{\arabic*}.}]
    \item Any two diagrams $g$ and $g'$ whose flat vertex graphs $\tilde G$ and $\tilde G'$ are isotopic have $R(g') = (-A)^n R(g)$ for some integer $n$.
    
    \item \label{iso:spatial}
    If every vertex has valence at most three, then two diagrams $g$ and $g'$ whose spatial graphs $\tilde G$ and $\tilde G'$ are isotopic have $R(g') = (-A)^n R(g)$ for some integer $n$.
    
    \item Any two diagrams $g$ and $g'$ whose associated ribbon graphs $\tilde G$ and $\tilde G'$ are isotopic have $R(g') = R(g)$.
\end{enumerate}

The next set of relations for the Yamada polynomial can be derived from the previous ones and are very useful aides for its calculation. Detailed proofs for these  relations (Y1-Y8) are provided in Ref.~\cite{Yamada1989}. They are as follows:

\begin{enumerate}[label={Y}{{\arabic*}.}]
      \item $R({\includegraphics[width = 0.4cm,  valign=c]{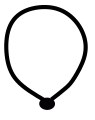}}) = B$, \quad where $B = A + 1 + A^{-1},$
      \item $R({\includegraphics[width = 0.8cm, valign=c]{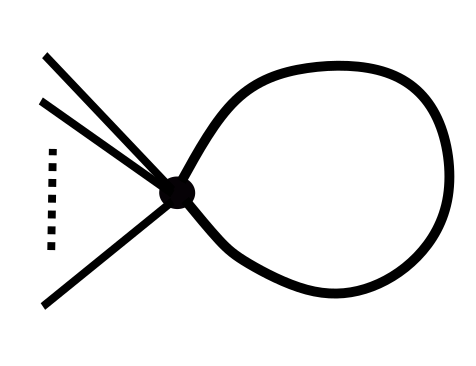}}) = -B R({\includegraphics[width = 0.6cm, valign=c]{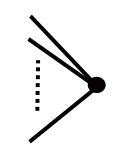}})$,
      \item $R({\includegraphics[width = 1.2cm, valign=c]{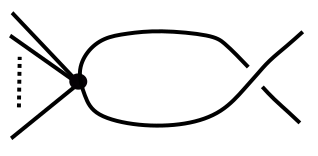}}) = -A R({\includegraphics[width = 0.8cm, valign=c]{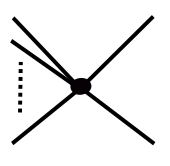}}) - (A^{2} + A) R({\includegraphics[width = 1cm, valign=c]{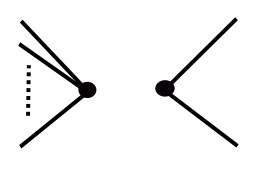}})$,
      \item $R({\includegraphics[width = 1.2cm, valign=c]{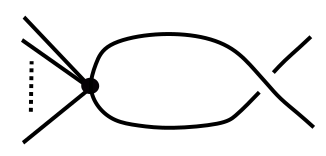}}) = -A^{-1}R({\includegraphics[width = 0.8cm, valign=c]{Figures/Eqn_fig12.PNG}}) - (A^{-2} + A^{-1}) R({\includegraphics[width = 1cm, valign = c]{Figures/Eqn_fig13.PNG}} )$,
      \item $R({\includegraphics[width = 1.2cm, valign=c]{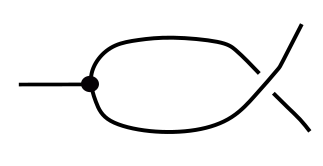}}) = -AR({\includegraphics[width = 1cm, valign=c]{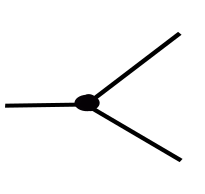}}),\text{ } R({\includegraphics[width = 1.2cm, valign=c]{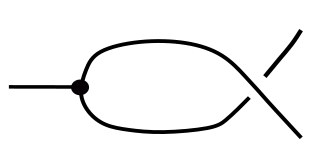}}) = -A^{-1} R({\includegraphics[width =1cm, valign=c]{Figures/Eqn_fig16.PNG}})$,
      \item $R({\includegraphics[width = 0.6cm, valign=c]{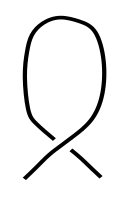}}) = A^{2} R({\includegraphics[width = 1cm, valign=c]{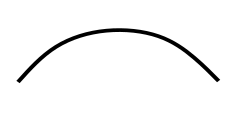}}),\text{ } R({\includegraphics[width = 0.6cm, valign=c]{Figures/Eqn_fig19.PNG}}) = A^{-2} R({\includegraphics[width = 1cm, valign=c]{Figures/Eqn_fig20.PNG}})$,
      \item $ \text{{Edge subdivision does not change the polynomial:}}\\ {\includegraphics[width = 4cm, valign=c]{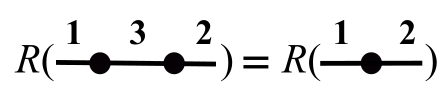}}$
     \item $ \text{Petals to concentric self-loops:}\\ {\includegraphics[width = 4cm, valign=c]{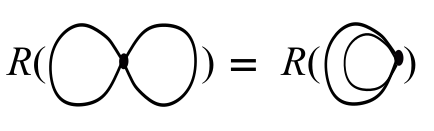}} = -B^{2} = -(A+ 1 A^{-1})^{2}$.
\end{enumerate}
\subsection{Illustrative Examples} \label{Examples}
 Yamada polynomials for a few spatial graphs are calculated by reducing the spatial graph diagram into a linear combination of smaller elements based on the skein relations stated above.\\
\textbf{Example 1 - Theta ($\theta_{1}$) graph}: The Yamada polynomial for a standard theta graph is calculated as follows:
\begin{align*}
R\Bigg({\includegraphics[width = 1cm , valign=c]{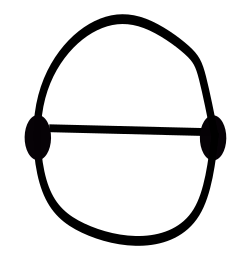}}\Bigg) = \hspace{-0.1cm} \underbrace{H\Bigg({\includegraphics[width = 1cm , valign=c]{Figures/Example_Fig_1.PNG}}\Bigg)}_{ \left(\parbox{1.75cm}{\scriptsize Apply S2 on the center edge}\right)\rightarrow} \hspace{-0.1cm}& = \underbrace{ H\Bigg({\includegraphics[width = 1cm , valign=c]{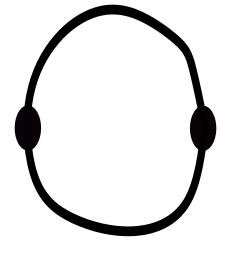}}\Bigg)}_{\downarrow (\text{Apply Y7})} + \underbrace{H\Bigg({\includegraphics[width = 2cm, valign=c]{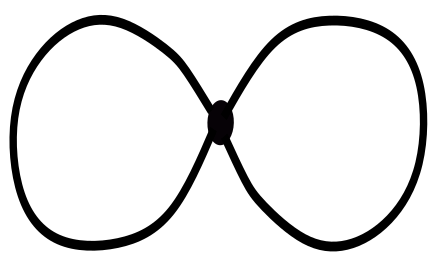}}\Bigg)}_{\downarrow (\text{Apply Y8})},\\ \implies R\Bigg({\includegraphics[width = 1cm , valign=c]{Figures/Example_Fig_1.PNG}}\Bigg) &= H\Bigg({\includegraphics[width = 1cm , valign=c]{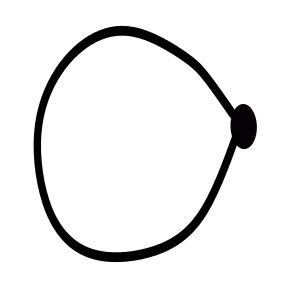}}\Bigg) + H\Bigg({\includegraphics[width = 1cm , valign=c]{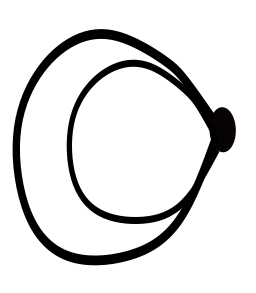}}\Bigg),\\ \implies
   R(\theta_{1}) &= B - B^{2}, \tag{where $B = A + 1 + A^{-1}$}\\\implies
   R(\theta_{1}) &= -(2+A+A^{-1}+A^{2}+A^{-2})
   \end{align*}

\textbf{Example 2:} A spatial graph  \Bigg({\includegraphics[width = 1cm , valign=c]{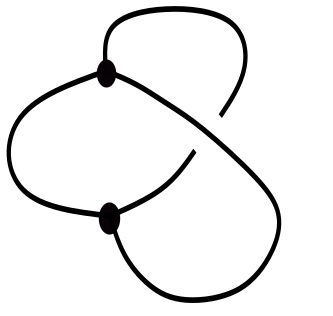}}\Bigg) which is isotopic (by the R6 move) to the standard theta graph. Its Yamada polynomial is calculated as follows, though one could instead use property Y5 as a shortcut.
\begin{align*}
   R\Bigg({\includegraphics[width = 1cm , valign=c]{Figures/Example_Fig_6.PNG}}\Bigg) &= AR\Bigg({\includegraphics[width = 1cm , valign=c]{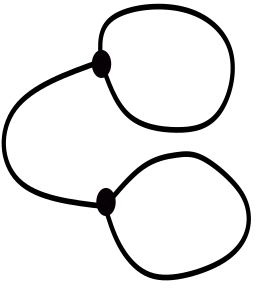}}\Bigg)  + A^{-1}R\Bigg({\includegraphics[width = 1cm , valign=c]{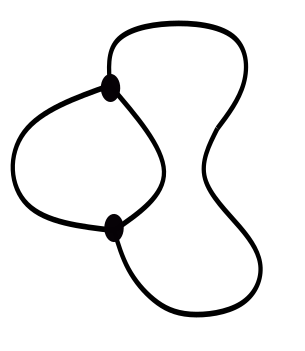}}\Bigg) + R\Bigg({\includegraphics[width = 1cm, valign=c]{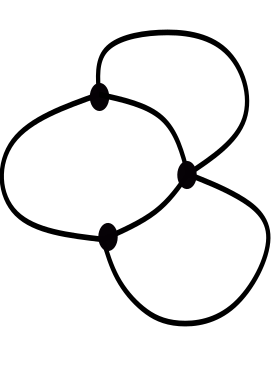}}\Bigg),\\  R\Bigg({\includegraphics[width = 1cm , valign=c]{Figures/Example_Fig_7.PNG}}\Bigg) &=0 \tag{because of isthmus based on property P3},\\
   R\Bigg({\includegraphics[width = 1cm , valign=c]{Figures/Example_Fig_1.PNG}}\Bigg) &=  H\Bigg({\includegraphics[width = 1cm , valign=c]{Figures/Example_Fig_1.PNG}}\Bigg) = H\Bigg({\includegraphics[width = 1cm , valign=c]{Figures/Example_Fig_2.PNG}}\Bigg) + H\Bigg({\includegraphics[width = 2cm, valign=c]{Figures/Example_Fig_3.PNG}}\Bigg),\\
   &= B-B^{2} \tag{where $B = A + 1 + A^{-1}$},\\
    R\Bigg({\includegraphics[width = 1cm , valign=c]{Figures/Example_Fig_9.PNG}}\Bigg) &=  H\Bigg({\includegraphics[width = 1cm , valign=c]{Figures/Example_Fig_9.PNG}}\Bigg) = H\Bigg({\includegraphics[width = 1cm , valign=c]{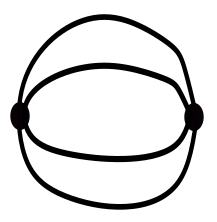}}\Bigg) + H\Bigg({\includegraphics[width = 1cm, valign=c]{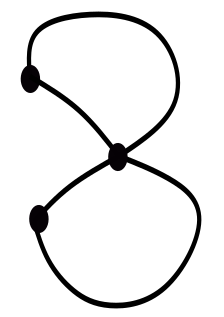}}\Bigg),\\
    &=  H\Bigg({\includegraphics[width = 1cm , valign=c]{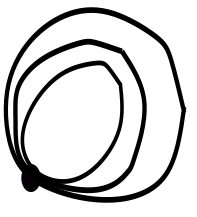}}\Bigg) + H\Bigg({\includegraphics[width = 1cm , valign=c]{Figures/Example_Fig_1.PNG}}\Bigg) + H\Bigg({\includegraphics[width = 2cm, valign=c]{Figures/Example_Fig_3.PNG}}\Bigg),\\
    &= B^{3} + (-B^{2} + B) + -B^{2} = B^{3}-2B^{2} + B,\\    \implies 
     R\Bigg({\includegraphics[width = 1cm , valign=c]{Figures/Example_Fig_6.PNG}}\Bigg) &= A(0) + A^{-1}(B-B^{2}) + (B^{3}-2B^{2} + B) , \\
     \implies 
     R\Bigg({\includegraphics[width = 1cm , valign=c]{Figures/Example_Fig_6.PNG}}\Bigg) &= A^{3} + A^{2} + 2A + 1 + A^{-1}, \\
     & = - A(R(\theta_{1})). \quad \text{Note the $-A$ factor permitted in \ref{iso:spatial}}\\
   \end{align*}
 
\textbf{Example 3:} The spatial graph is \Bigg({\includegraphics[width = 1cm , valign=c]{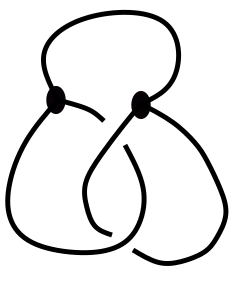}}\Bigg). This example involves extensive use of Yamada polynomial skein relations (Y1-Y8). Its Yamada polynomial is calculated as follows: 
\begin{align*}
   &= \underbrace{AR\Bigg({\includegraphics[width = 1cm , valign=c]{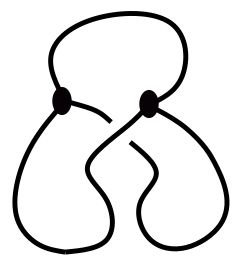}}\Bigg)}_{\downarrow (\text{Apply Y5})}  + \underbrace{A^{-1}R\Bigg({\includegraphics[width = 1cm , valign=c]{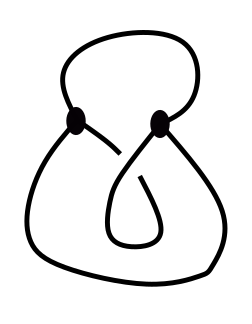}}\Bigg)}_{\downarrow (\text{Apply Y6})} + R\Bigg({\includegraphics[width = 1cm, valign=c]{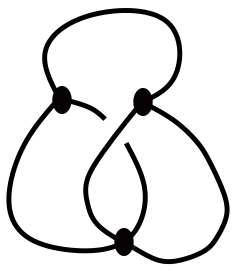}}\Bigg), \\
   & \text{Drop the Rs for simplicity},\\
   &= A(-A \Bigg({\includegraphics[width = 1cm , valign=c]{Figures/Example_Fig_1.PNG}}\Bigg) + A^{-1}(A^{-2} \Bigg({\includegraphics[width = 1cm , valign=c]{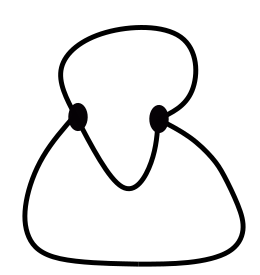}}\Bigg)) + \underbrace{\Bigg({\includegraphics[width = 1cm , valign=c]{Figures/Example_Fig_16.PNG}}\Bigg)}_{\downarrow (\text{Apply Y4})},\\
   &= (-A^{2} + A^{-3}){\includegraphics[width = 1cm , valign=c]{Figures/Example_Fig_1.PNG}} + -A^{-1} {\includegraphics[width = 1cm , valign=c]{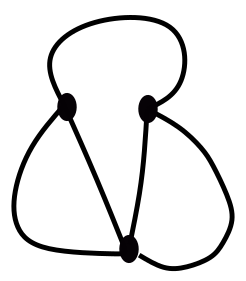}} -(A^{-2}+ A^{-1}){\includegraphics[width = 1cm , valign=c]{Figures/Example_Fig_17.PNG}}     \\
   &= (-A^{2}-A^{-2}- A^{-1} + A^{-3}){\includegraphics[width = 1cm , valign=c]{Figures/Example_Fig_1.PNG}} - A^{-1}{\includegraphics[width = 1cm , valign=c]{Figures/Example_Fig_18.PNG}},\\
  &=  A^4 + A^3 + A^2 + A - A^{-2} - A^{-3} - A^{-4} - A^{-5} \\
  & \text{Since taking $B = A + 1 + A^{-1}$ we have:}\\
  {\includegraphics[width = 1cm , valign=c]{Figures/Example_Fig_18.PNG}} &=  {\includegraphics[width = 1cm , angle= 90, valign=c]{Figures/Example_Fig_10.PNG}} +  {\includegraphics[width = 1cm , angle= -90, valign=c]{Figures/Example_Fig_11.PNG}} = {\includegraphics[width = 1cm ,  valign=c]{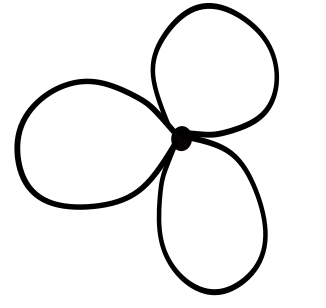}} + {\includegraphics[width = 1cm , angle= 90, valign=c]{Figures/Example_Fig_1.PNG}}  + {\includegraphics[width = 1.5cm , valign=c]{Figures/Example_Fig_3.PNG}},\\
  & =  + B^{3} + (-B^{2}+B) + - B^{2} = B^{3} - 2B^{2} + B.
  \end{align*}  

\textbf{Example 4:} The spatial graph is \Bigg({\includegraphics[width = 1.2cm , valign=c]{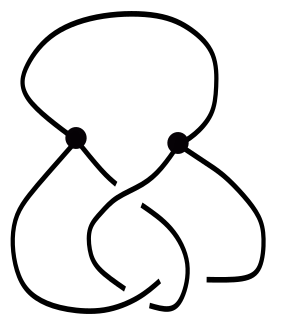}}\Bigg). This example involves extensive use of skein relations (Y1-Y8). Its Yamada polynomial is calculated as follows: 
\begin{align*}
  {\includegraphics[width = 1.2cm , valign=c]{Figures/Example_Fig_20.PNG}}&= AR\Bigg({\includegraphics[width = 1cm , valign=c]{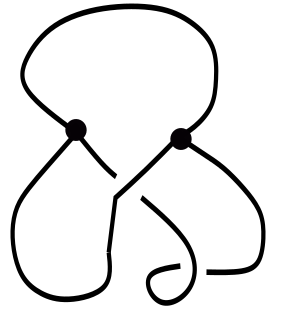}}\Bigg)  + A^{-1}R\Bigg({\includegraphics[width = 1cm , valign=c]{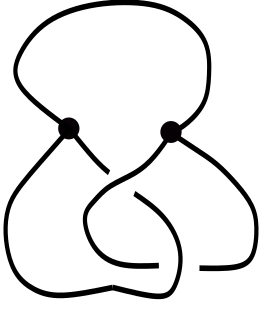}}\Bigg) + R\Bigg({\includegraphics[width = 1cm, valign=c]{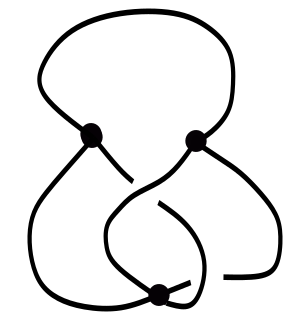}}\Bigg),\\
   & \text{Drop the Rs for simplicity},\\
   &= A^{3}\underbrace{{\includegraphics[width = 1cm , valign=c]{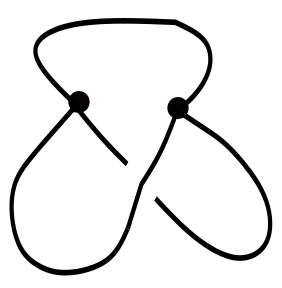}}}_\text{Example 2} + A^{-1}\underbrace{{\includegraphics[width = 1cm , valign=c]{Figures/Example_Fig_13.PNG}}}_\text{Example 3} - A {\includegraphics[width = 1cm , valign=c]{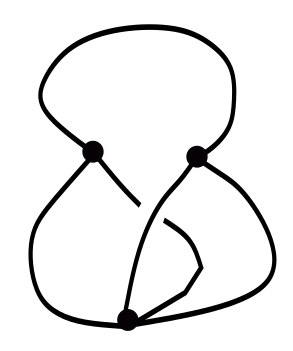}} -(A^{2} + A){\includegraphics[width = 1cm , valign=c]{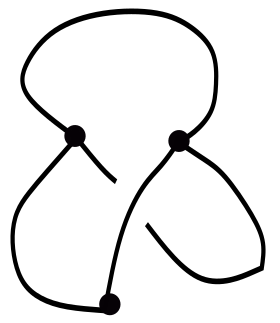}},\\
   &= -A^{4}{\includegraphics[width = 1cm , valign=c]{Figures/Example_Fig_1.PNG}} + -A^{-1} {\includegraphics[width = 1cm , valign=c]{Figures/Example_Fig_13.PNG}}  - A(-A^{-1}) {\includegraphics[width = 1cm , valign=c]{Figures/Example_Fig_18.PNG}} \\ & -(A^{-2}+ A^{-1}){\includegraphics[width = 1cm , valign=c]{Figures/Example_Fig_17.PNG}} - (A^2 + A)(-A){\includegraphics[width = 1cm , valign=c]{Figures/Example_Fig_1.PNG}}      \\
  &=  (-A^4 + A^{-1} + 1 + A^3 + A^2) {\includegraphics[width = 1cm , valign=c]{Figures/Example_Fig_1.PNG}}  +  A^{-1}{\includegraphics[width = 1cm , valign=c]{Figures/Example_Fig_13.PNG}}  + {\includegraphics[width = 1cm , valign=c]{Figures/Example_Fig_18.PNG}}  \\
  & =  A^{6}-A^{2}-1 - A^{-2} - A^{-3}  - A^{-4}  - A^{-5}  - A^{-6}.
  \end{align*}

\section{PROPOSED SGD ENUMERATION FRAMEWORK} \label{Framework}
Figure~\ref{Flowchart} shows the steps of the proposed design framework to represent, enumerate and categorize unique spatial topologies of a 3D system (assuming inter-component connectivity is fixed). The detailed steps of the enumeration design framework are: 
\begin{itemize}
    \item \textbf{1. Define system architecture:} Provide the specific 3D system architecture (SA) for which spatial topologies must be enumerated. From the SA, extract the number of  nodes (components), their valencies, system interconnectivity, and the corresponding edges (interconnects in the system). 
    \item \textbf{2. Enumerate spatial graph diagrams:} Combinatorially enumerate all possible spatial graph diagrams (SGDs) for the SA from zero crossings to the maximum crossing number ($k= 0, 1, ..., k_{m}$) using their corresponding shadows.
    \item \textbf{3. Check graph planarity:} Planar diagrams (PDs) of spatial graphs are used for the calculation of the Yamada polynomials.  However, before calculation, each enumerated graph must be checked to determine whether it is planar, for which there are linear-time algorithms \cite{HopcroftTarjan}.  The graphs start with a circular order of the edges at each vertex, making the planarity check even easier.  The algorithm shown in Fig.~\ref{PlanarityCheck} recursively contracts the edges of a graph until the diagram is a bouquet of circles, and then use the fact that if the diagram is planar, there must exist at least one loop edge whose endpoints come consecutively in the cyclic ordering around the vertex, i.e.,~a self-loop. This self-loop can be removed without altering planarity. 
    The recursive steps for the planarity check (PC) algorithm are enumerated below:
    \begin{enumerate}[label={PC}{{\arabic*}}.]
        \item Convert all vertices to crossings, as it does not affect planarity.
        \item Contract all non-loop edges (edges shared between two vertices) to a vertex, as it does not affect planarity.
        \item Remove all planar self-loops at a vertex.
         \item Empty vertex does not affect planarity, so remove it.
         By doing all these steps recursively, if the result is an empty diagram, then the original diagram is planar. If not, the diagram is non-planar. 
    \end{enumerate}
\begin{figure}
		\centering
		\includegraphics[width=3.3in]{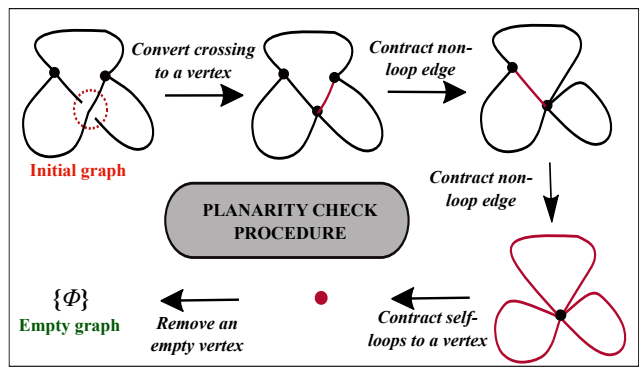}
		\caption{Identifying connected planar spatial diagrams from the combinatorially enumerated set is performed using this procedure.}
		\label{PlanarityCheck}
\end{figure}

 \begin{figure}
		\centering
		\includegraphics[width=4.7in]{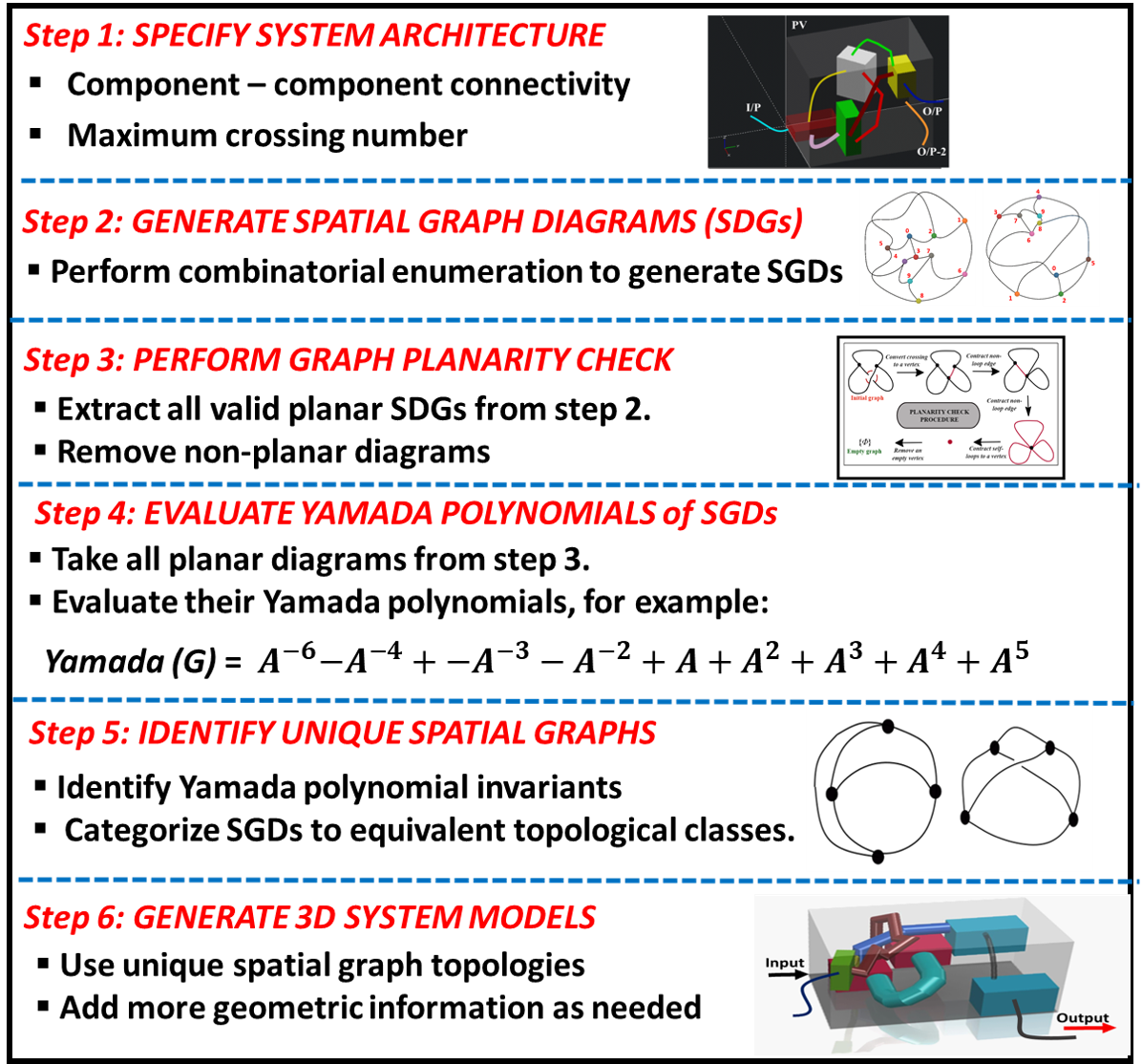}
		\caption{The six sequential steps of the proposed spatial graph-based topology enumeration framework.}
		\label{Flowchart}
\end{figure}
    
    \item \textbf{4. Evaluate Yamada polynomials:} The Yamada polynomials for all the valid planar SGDs are evaluated using the Yamada polynomial properties detailed in Sec.~\ref{Yamada}. 
    \item \textbf{5. Categorize different spatial topologies}:  Cluster SGDs into topological classes so that the SGDs belonging to each class have the same Yamada polynomial or differ by a factor $(-A)^{n}$ based on property ~\ref{iso:spatial} Consequently, no two classes have the same Yamada polynomial. 3D geometric models can then be generated for SGDs under different spatial topology classes as required.
    \item \textbf{6. 3D model generation:} SGDs from step 5 are utilized as underlying skeleton structures for generating various 3D system geometric models. As discussed in step 5, any pair of SGDs that represent isotopic spatial graphs will be in the same Yamada class. 3D geometric models for SGDs within the same topology class can be generated based on two different design requirements as follows:
    \begin{enumerate}
        \item In some engineering applications, additional crossings beyond what is required between a pair of interconnects may be undesirable. For example, there may be no advantage to two wires intertwining several times unless they intentionally function as a twisted pair.  Similarly, crossings between pair of pipes may be difficult to fabricate, and more complex to install and remove. Having fewer crossings helps reduce complexity and helps in easier system diagnosis and maintenance. In such cases, an SGD having the fewest crossings is selected to generate a 3D geometric model. 
        \item However, in some applications, this assumption that the simplest SGD is desirable may not always be applicable. Sometimes due to interface requirements, interconnects may require to exhibit complex topology (e.g., pipes or electrical lines passing through a fire-rated interface in a wall). Other examples include cases where multiple crossings enable easier assembly, disassembly, design effective wire harnesses, or simpler instantiating of other SGD elements. Quite often a collection of intertwining wires or ducts can help increase mechanical strength to provide support and conserve space.
    \end{enumerate}
\end{itemize}

\section{CASE STUDIES} \label{CaseStudy}
In this section, a number of case studies are provided to demonstrate the proposed enumeration framework discussed in Sec.~\ref{Framework}. All computations (Yamada polynomial calculation, planarity check, etc.) in the case studies were performed using \textsc{Wolfram Mathematica 11.3} software  with an Intel Xeon E5-2660 CPU @ 2.00 GHz, 64 GB DDR4-2400 RAM, \textsc{Windows} 10 64-bit workstation.

\subsection{Case Study 1: Components with equal valencies} \label{CS1}
\begin{figure}
		\centering
		\includegraphics[width=5in]{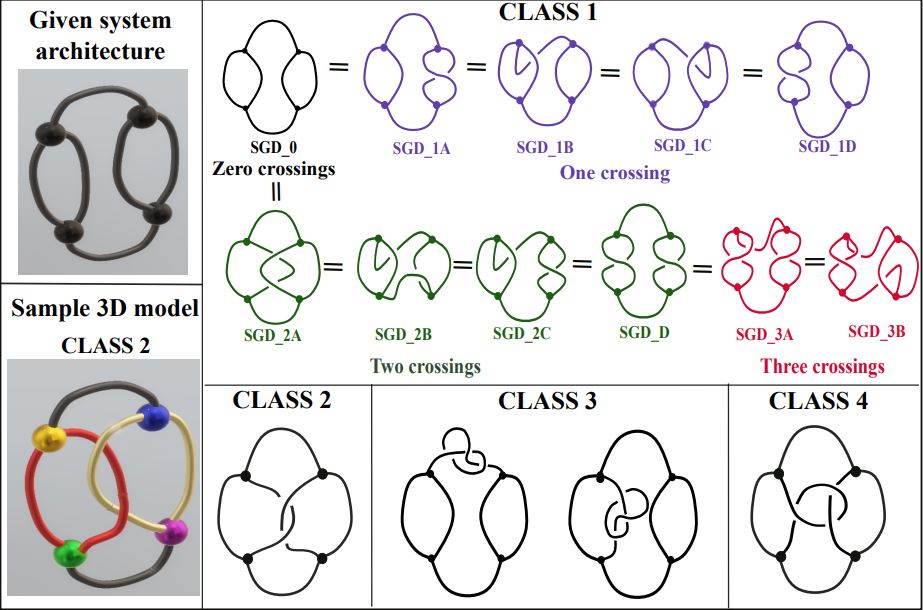}
		\caption{Results from case study~\ref{CS1} for a given system architecture for spatial graphs with interconnect crossing numbers from 0 through 3.}
		\label{CaseStudy1}
\end{figure}

In this case study, we consider a 3D system with architecture as shown in Fig.~\ref{CaseStudy1}. This system contains four identical trivalent components (nodes),  and six interconnects (edges). We find the unique 3D spatial topologies (STs) of the system for crossing numbers varying from zero to three. The notation used to indicate each SGD is given by SGD\_k where $k$ is the crossing number of that diagram, and the letter refers to the specific SGD. SGD\_0 is the original system architecture without any crossings. Using the proposed framework, we combinatorially enumerate all the SGDs and pass them through a planarity check procedure. Yamada polynomials were then calculated for 3, 31, 118, and 231 valid planar SGDs having 0, 1, 2 and 3 crossing numbers respectively. We group the SGDs having the same Yamada polynomial or differing by a factor $(-A)^{n}$ under the same topological class based on property ~\ref{iso:spatial}. Through this we attain a total of four unique Yamada classes as shown in Fig.~\ref{CaseStudy1}. The Yamada polynomials for these different classes of topologies are shown in Table~\ref{Table 1}. For class 1, it is observed that three distinct Yamada polynomials exist for different crossing numbers, but they all differ by a factor $(-A)^{n}$.  This strongly suggests that the SGDs shown under class 1 are isotopic to SGD\_0, and indeed this can be verified using the R6 move. In contrast, the two SGDs in class 3 can be shown to be nonisotopic using other tools.  A sample 3D model of class 2 spatial topology candidate design solution is also shown in Fig.~\ref{CaseStudy1}. The computational time for this entire case study is 78.3 seconds (s).

\begin{table}[htbp]
\caption{Yamada polynomials of diagrams shown in Fig.~\ref{CaseStudy1} (case study~\ref{CS1}).}
\label{Table 1}
\begin{center}
  \footnotesize\addtolength{\tabcolsep}{-5pt}
\begin{tabular}{c c c c} 
 \hline\hline
 \textbf{Classes}   &  \textbf{Yamada polynomials} \\ [0.5ex]
 \hline
Class 1    SGD\_0: & $A^{-3} + A^{-2} + 3A^{-1}+ 2 + 3A + A^{2} + A^{3} $ \\ 

 Class 1    SGD\_1: & $(-A)^{1}(A^{-3} + A^{-2} + 3A^{-1}+ 2 + 3A + A^{2} + A^{3})$ \\ 
Class 1    SGD\_2:  & $(-A)^{2}(A^{-3} + A^{-2} + 3A^{-1}+ 2 + 3A + A^{2} + A^{3})$ \\ 
 Class 1   SGD\_3:  & $(-A)^{3}(A^{-3} + A^{-2} + 3A^{-1}+ 2 + 3A + A^{2} + A^{3})$ \\ 
\hline
Class 2   & $A^{-4} + A^{-3} + A^{-2} + 2A^{-2} + 2A + 2A^{3} + A^{4} + A^{5} + A^{6}$ \\ 
\hline
Class 3  & $A^{-7} + A^{-6} + 3A^{-5} + 3A^{-4} +4A^{-3} +3A^{-2} +3A^{-1}$ \\  &  $-3A^{2} -2A^{3} -2A^{4} -A^{5} +A^{6} +A^{8}$ \\
\hline
Class 4  & $A^{-7} + A^{-6} + A^{-5} + 2A^{-4} +2A^{-3} +A^{-2} +3A^{-1} + 2A -A^{2} -A^{5} +A^{6}$ \\
 \hline\hline
\end{tabular}
\end{center}
\end{table}

\subsection{Case Study 2: Enumeration of Spatial Topologies of an Automotive Fuel Cell System (AFCS)}\label{CS2} 
There are two main goals of this case study. First, to demonstrate the proposed spatial topology enumeration framework on a practical industry application, the Ford automotive 3D fuel cell system (AFCS). Second, to  provide insight into how the number of unique spatial topologies varies with increasing the number of components and the maximum number of crossings in a 3D system.\\

\begin{figure}
\centering
	\begin{subfigure}{0.55\linewidth}
		\includegraphics[width=1\linewidth]{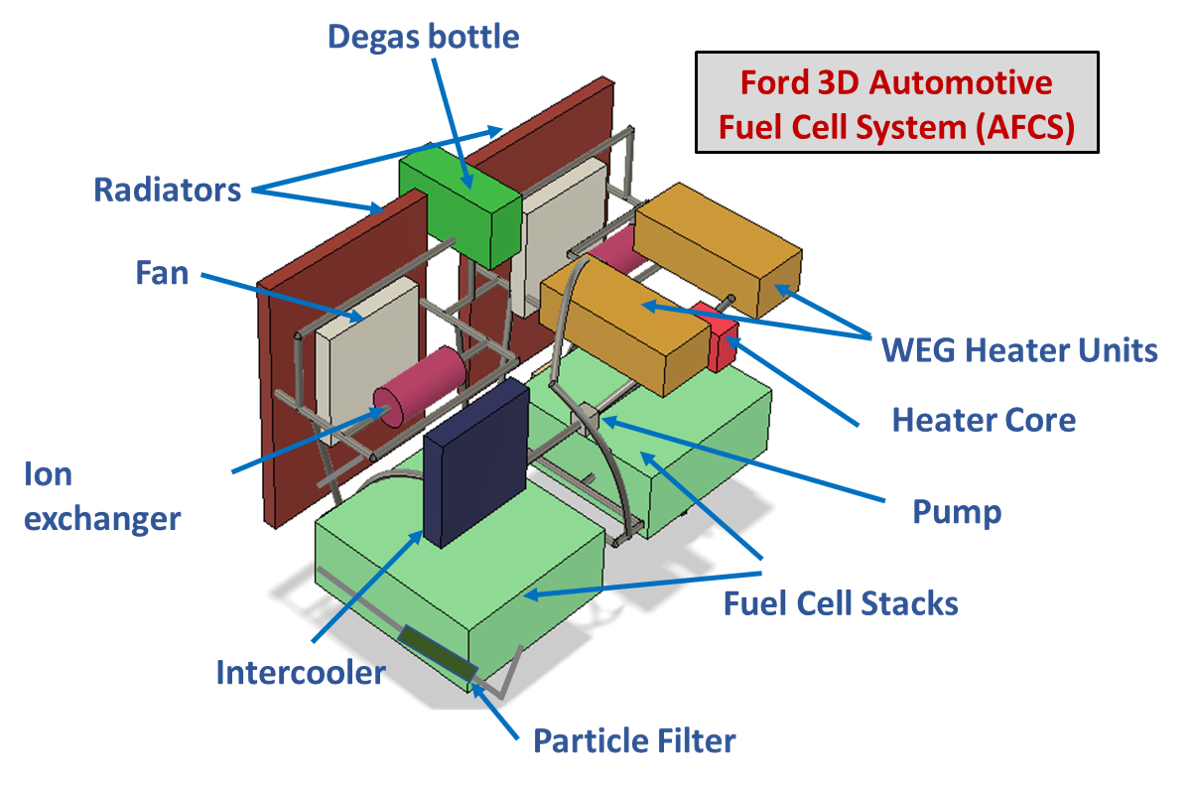}
		\caption{}
		\label{3D_AFCS_Layout}
	\end{subfigure}
	\begin{subfigure}{0.44\linewidth}
\includegraphics[width=1\linewidth]{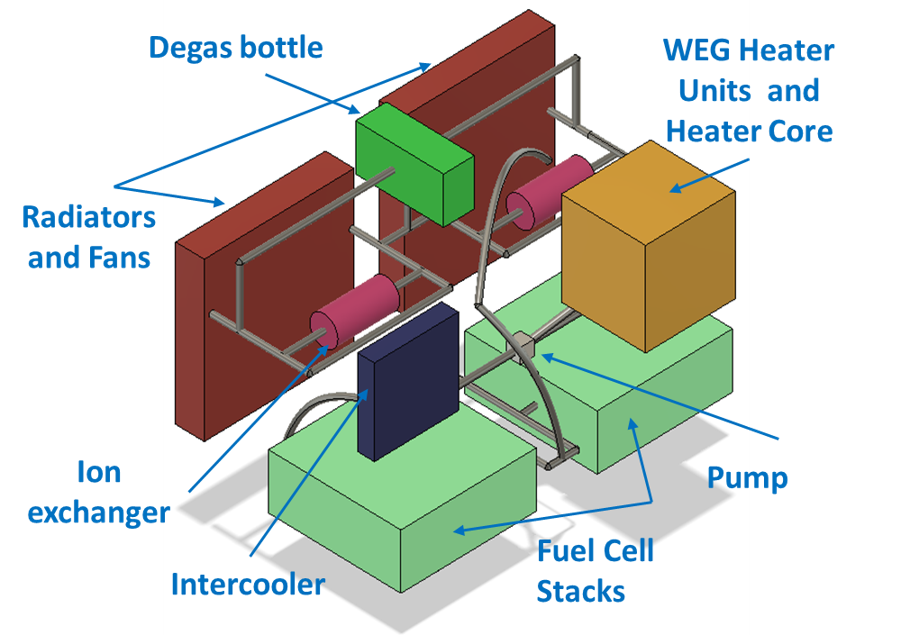}
		\caption{}
		\label{AFCS_10_Component}
	\end{subfigure}
	\caption{a) An original 14-component layout of the 3D Automotive Fuel Cell System (AFCS) with its components and the interconnect network; b) A 10-component, simplified 3D layout of the AFCS used in Case Study~\ref{CS2} as one of the system architectures. In the Fig.~\ref{AFCS_10_Component} layout, some of the closest components from the 14-component layout (Fig.~\ref{3D_AFCS_Layout}) are combined together to a form a larger component. For instance, the two WEG heater units and heater core in  Fig.~\ref{3D_AFCS_Layout} are combined together into a single larger component in Fig.\ref{AFCS_10_Component}.}
	\label{AFCSFig}
\end{figure}

\textbf{Automotive Fuel Cell System (AFCS) Description:}  An AFC system contains fuel cell stacks, the required cooling system components  such as heat exchangers, pumps, radiators, cooling fans, particle filters, cabin heat generation unit, etc., and a cooling hose interconnect network as shown in Fig.~\ref{3D_AFCS_Layout}. Underhood 3D spatial packaging of the AFCS is essential for efficient thermal management in FCEVs (Fuel Cell Electric Vehicles). However, this requires finding a suitable AFCS 3D spatial topology that can be optimized to minimize underhood volume while delivering the required vehicle capability and physics-based system performance. Unique AFCS spatial topologies can be obtained using the proposed enumeration framework to thoroughly navigate through the discrete 3D design space for a given AFCS architecture. These unique spatial configurations can be utilized as initial 3D designs to perform parametric gradient-based multiphysics optimization to reduce overall AFC system volume subject to geometric and physics-based constraints. Finally, an AFCS spatial topology configuration is selected that satisfies all the performance criteria and constraints. However, this optimization process is beyond the scope of this article and has been separately addressed in detail in our previous articles~\cite{Bhattacharyya2022_AIAA, Jessee2020a, Peddada2020, Peddada2020b, Peddada2020_JMD_2Stage, Peddada2021PhD}. In this study we are only interested in demonstrating an important use case scenario of this proposed enumeration framework.\\

Here we consider only AFCS system architectures where all nodes (components) are trivalent. The original AFCS system shown in Fig.~\ref{3D_AFCS_Layout} contains 14 trivalent components. To test our framework on \textit{highly connected} AFCS architectures containing different number of components, we group together some of the closer or smaller components of the 14-component 3D AFCS layout to create architectures having 6, 8, 10, and 12 components. For example, Fig.~\ref{AFCS_10_Component} shows an AFCS architecture containing only 10 trivalent components where the two WEG heater units and heater core are combined to a single larger component, the small particle filter is merged into the full cell stack, and the two cooling fans are combined with the radiators, thus reducing a 14-component system to a 10-component system architecture. Then, unique spatial topologies are enumerated for such AFCS system architectures containing 6, 8, 10, 12, and 14 trivalent components with interconnect crossing numbers varying between 0 through 6. Table~\ref{Table 4} shows the number of topologically distinct AFCS spatial configurations (with distinct Yamada polynomials) generated by this study.\\

\textbf{Running times:} Tables~\ref{Table 2} and \ref{Table 3} show the enumeration running times in seconds, and in log-base-10, respectively, for different system architectures with varying number of nodes and crossings in the spatial graph diagrams. Computations were performed in parallel on a 10-core Xeon workstation using code written in Python. Note that increasing either the number of nodes by 2 or the number of crossings by 1 increases the running time by a factor of 10.\\

Figure~\ref{AFCS_Study_Topologies} shows a sampling (6 out of 37) of the 3D geometric models that have been automatically generated for a 10-component system with 3 crossings. Similarly, Fig.\ref{AFCS_Study_Topologies2} shows 3 out of 283 spatial topologies for a 10-component system with 4 interconnect crossings.  The top and orthogonal views along with their corresponding spatial graphs were generated using Autodesk Fusion 360 software. The top view helps in visualizing the component-component connectivity and the crossings between the interconnects. Please note that the components and interconnects in Fig.~\ref{AFCS_Study_Topologies} are laid out much farther apart from each other compared to the original AFCS layout in Fig.~\ref{AFCS_10_Component} to help visualize the different spatial diagrams clearly. Figure~\ref{AFCS_Study_Topologies3} shows two out of   69 possible unique spatial topologies of the full-scale 3D automotive fuel cell system (AFCS) shown in Fig.~\ref{3D_AFCS_Layout} containing 14 components and 3 crossings.

\begin{table}[htbp]
\caption{Running times (in seconds) to generate unique spatial topology classes of AFCS with varying number of components and crossings in Case Study~\ref{CS2}.}
\label{Table 2}
\begin{center}
   \small\addtolength{\tabcolsep}{0pt}
 \begin{tabular}{c c l l l l l l} 
 \hline\hline
 \diagbox[dir=NW]{\textbf{Components}}{\textbf{Crossings}} & 0 &  1 & 2 & 3 & 4 & 5 & 6\\ [0.5ex]
 \hline
 6 &  0.03 &  0.06 & 0.11 & 1.03 & 9.46 & 98.44 & 1103.10 \\ 
 \hline
 8 & 0.04 &  0.09 & 0.89 & 14.35 & 224.73 & 3087.12 &  \\
 \hline
 10 & 0.05 &  0.37 & 9.26 & 213.36 & 4202.83 &  &  \\
 \hline
 12 & 0.10 &  2.37 & 98.08 & 2879.99 &  &  &   \\
\hline
14 & 0.40 &  25.17 & 1258.59 & 45363.99 &  &  &  \\
 \hline\hline
\end{tabular}
\end{center}
\end{table}

\begin{table}[htbp]
\caption{Running times of Case Study~\ref{CS2} (in log--base-10 seconds) to generate unique spatial topology classes with increasing number of components and crossings.}
\label{Table 3}
\begin{center}
\small\addtolength{\tabcolsep}{0pt}
 \begin{tabular}{c c c c c c c c c} 
 \hline\hline
 \diagbox[dir=NW]{\textbf{Components}}{\textbf{Crossings}} & 0 &  1 & 2 & 3 & 4 & 5 & 6\\ [0.5ex]
 \hline
 6 &  -1.47 &  -1.25 & -0.96 & 0.01 & 0.98 & 1.99 & 3.04 \\ 
 \hline
 8 & -1.43 & -1.06 & -0.05 & 1.16 & 2.35 & 3.49 &  \\
 \hline
 10 & -1.30 &  -0.44 & 0.97 & 2.33 & 3.62 &  &  \\
 \hline
 12 & -1.01 &  0.37 & 1.99 & 3.46 &  &  &   \\
\hline
14 & -0.39 &  1.40 & 3.10 & 4.66 &  &  &  \\

 \hline\hline
\end{tabular}
\end{center}
\end{table}

\begin{table}[htbp]
\caption{Number of unique spatial topology classes with increasing number of components and crossings.}
\label{Table 4}
\begin{center}
\small\addtolength{\tabcolsep}{0pt}
 \begin{tabular}{c c c c c c c c c} 
 \hline\hline
 \diagbox[dir=NW]{\textbf{Components}}{\textbf{Crossings}} & 0 &  1 & 2 & 3 & 4 & 5 & 6\\ [0.5ex]
 \hline
 6 &  0 &  1 & 0 & 3 & 5 & 35 & 211 \\ 
 \hline
 8 & 0 & 1 & 1 & 7 & 36 & 294 &  \\
 \hline
 10 & 0 & 1 & 3 & 37 & 283 &  &  \\
 \hline
 12 & 0  & 1 & 5 & 74 &  &  &   \\
\hline
14 & 0 &  0 & 5 & 69 &  &  &  \\
 \hline\hline
\end{tabular}
\end{center}
\end{table}

\subsection{Case Study 3: Components with different valencies}\label{CS3}
\begin{figure}
		\centering
		\includegraphics[width=4.5in]{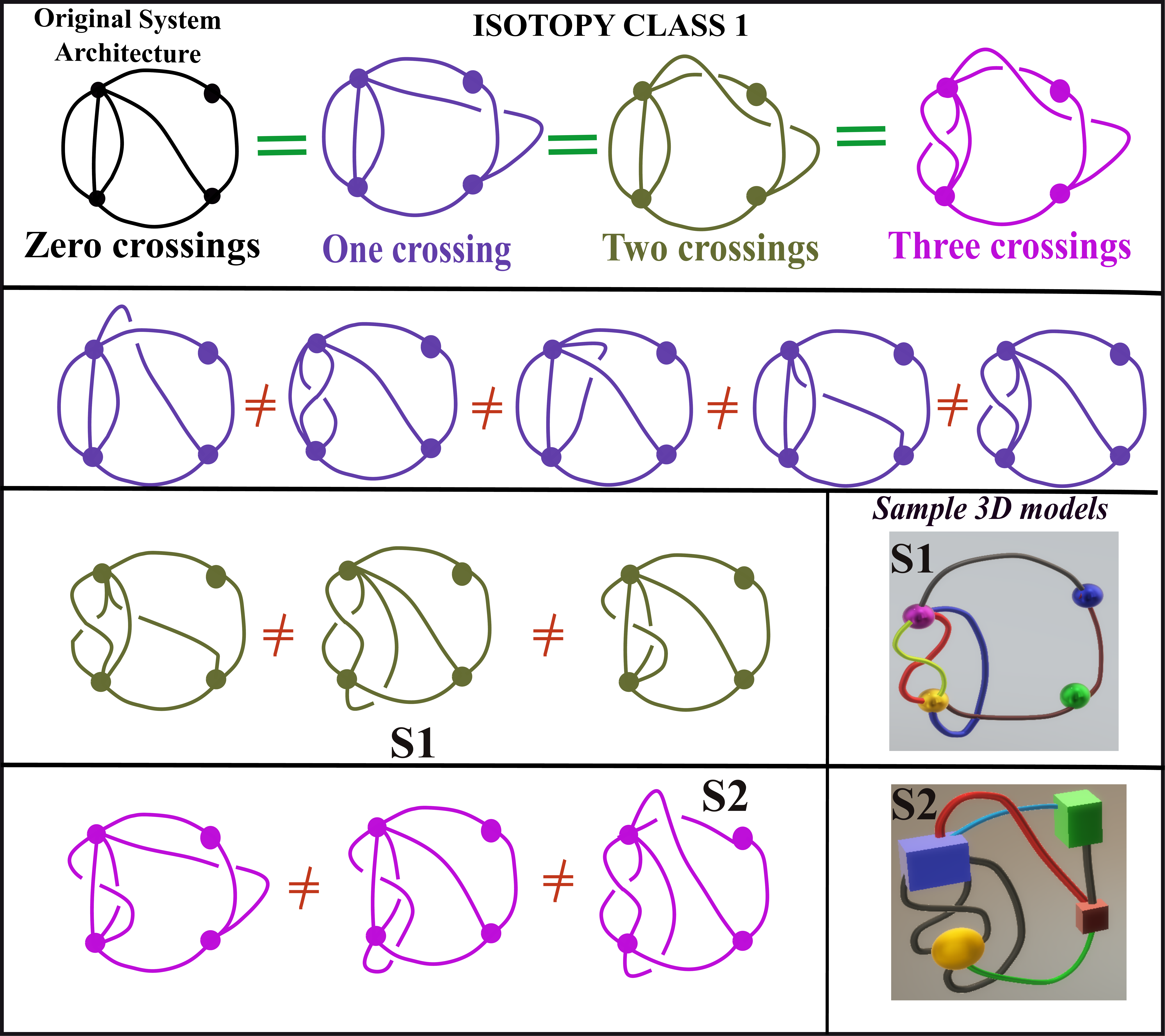}
		\caption{Results from case study~\ref{CS3} for a given system architecture for maximum crossing numbers from 0 through 3 with components having different valencies.}
		\label{Casestudy3}
\end{figure}

\begin{figure}
	\centering
		\includegraphics[width=4.6in]{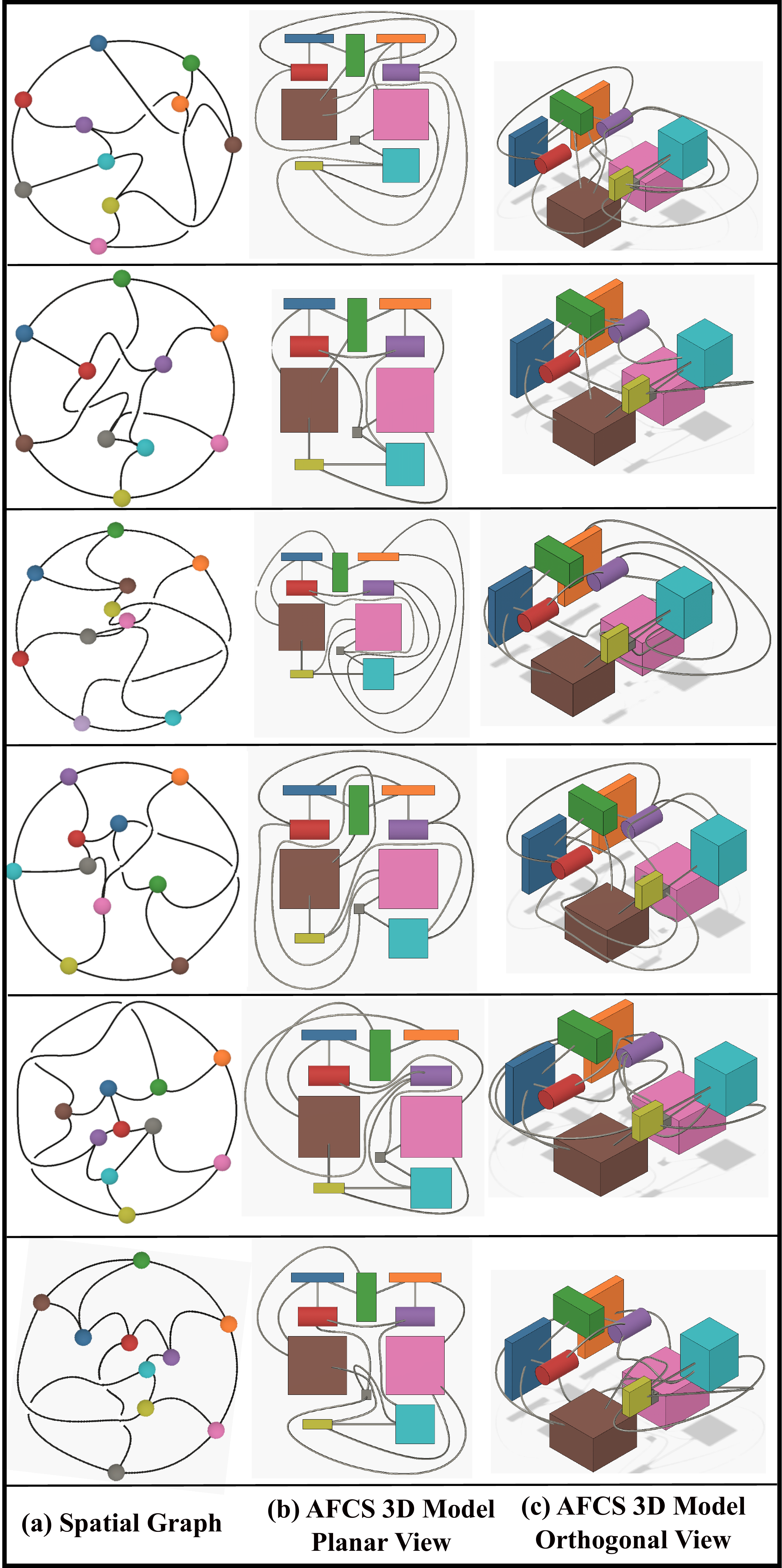}
		\caption{Six unique spatial topologies of the automotive fuel cell system  discussed in case study~\ref{CS2}.}
		\label{AFCS_Study_Topologies}
\end{figure}

\begin{figure}
	\centering
		\includegraphics[width=5in]{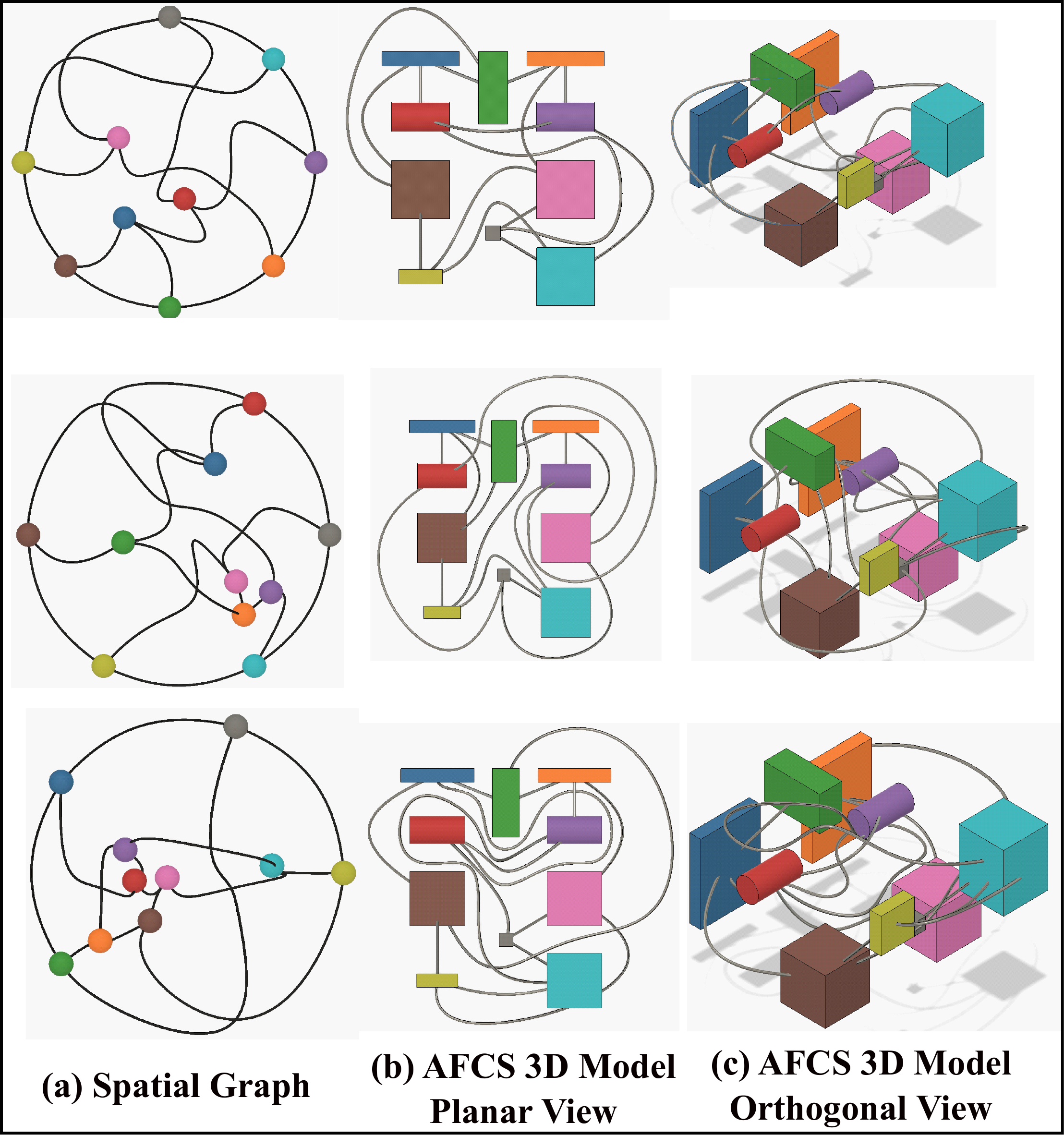}
		\caption{Three unique spatial topologies of the automotive fuel cell system  with 10 components and 4 crossings discussed in case study~\ref{CS2}.}
		\label{AFCS_Study_Topologies2}
\end{figure}

\begin{figure}
	\centering
		\includegraphics[width=6.5in]{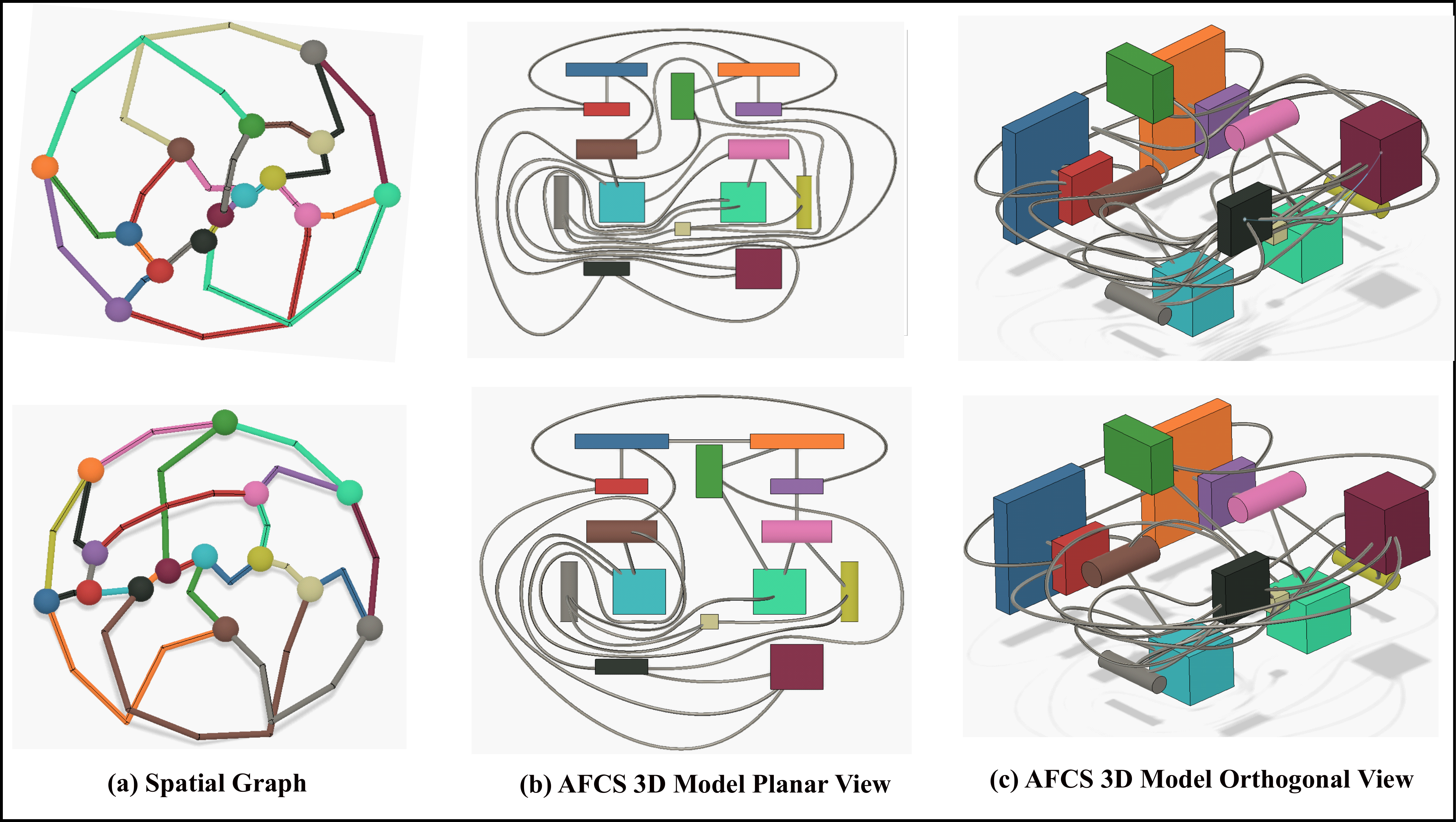}
		\caption{From Case Study~\ref{CS2}, here are two unique spatial topologies of the full-scale 3D automotive fuel cell system (AFCS) shown in Fig.~\ref{3D_AFCS_Layout} containing 14 components and 3 crossings.}
		\label{AFCS_Study_Topologies3}
\end{figure}

In most real-world systems, every component may not have the same valency (number of ports). Unlike case study~\ref{CS1}, in this case study we investigate a system with four components having different valencies. In addition, flat vertex graphs (FVGs) as described in Sec.~\ref{FlatVertex} are used. As FVGs have local structures at nodes, the edge connectivity order around the nodes is preserved, and thus FVG representations are highly suitable for design applications where nodes have a specific cyclic ordering of ports. Here R0 to R5 moves are valid but not R6. Figure~\ref{Casestudy3} shows some of the results obtained in this study. After computing the Yamada polynomials of hundreds of planar SGDs, a total of 27 unique Yamada classes are obtained. For illustration purposes, we show some isotopes of SGD\_0 (original  system architecture) as class 1 isotopes. Furthermore, unique SGDs belonging to some unique Yamada classes are shown for crossing numbers one, two, and three respectively. Two final 3D system geometric models (referred as S1 and S2) are also shown in Fig.~\ref{Casestudy3}. The total computational time taken for study B was 211.4 sec. It can be observed from this study that with components with different valencies, we get more unique Yamada classes than those with identical components. Thus, manually generating such designs is very challenging and the automated enumeration framework we proposed here is very valuable. \\ 

{\begin{table}[htbp]
\caption{Yamada polynomials of diagrams shown in Fig.~\ref{AFCS_Study_Topologies} (case study~\ref{CS2}).}
\label{Table 5}
\begin{center}
  \footnotesize\addtolength{\tabcolsep}{-5pt}
\begin{tabular}{c c c c} 
 \hline\hline
 \textbf{Classes}   &  \textbf{Yamada polynomials} \\ [0.5ex]
 \hline
Class 1:  & $-1 - A + 2A^2 - 3A^3 - 2A^4 + 13A^5 - 24A^6 + 3*A^7 - 39A^8 $ \\  & $ + 42A^9 - 30A^{10} + 23A^{11} - 5A^{12} + 7A^{14} - 3A^{15} + 2A^{16}$\\
\hline
Class 2 : & $2 - 4A + 2A^2 + 2A^3 - 18A^4 + 24A^5 - 41A^6 + 37A^7$ \\  & $ - 41A^8 + 28A^9 - 18A^{10} + 6A^{11} + 3A^{12} - 3A^{13} + 3A^{14}$\\
\hline
Class 3:  & $-1 - A^2 - 2A^3 - 3A^4 + 6A^5 - 19A^6 + 26A^7 - 36A^8  + 36A^9 $ \\  & $ - 35A^{10} + 25A^{11} - 16A^{12} + 4A^{13} + A^{14} - 4A^{15} + 2A^{16} - A^{17}$\\ 
\hline
Class 4:   & $ -1 + 2A - 5A^2 - A^3 - 6A^5 - 8A^6 + 18A^7 - 36A^{8} + 45A^{9} - 47A^{10} + $ \\  & $42A^{11} - 28A^{12} + 10A^{13} + 5A^{14} - 14A^{15} + 11A^{16} - 6A^{17} + A^{18}$ \\ 
\hline
Class 5: & $-2 - A^2 - 4A^3 + 7A^5 - 19A^6 + 33A^7 - 40A^8 + 42A^9 - 39A^{10}  $ \\  & $ + 24A^{11} - 15A^{12} - 3A^{13} + 5A^{14} - 8A^{15} + 3A^{16} - A^{17}$ \\ \hline
Class 6:  & $-1 + 3A - 6A^2 + 4A^3 - 2A^4 - 3A^5 - 5A^6 + 14A^7 - 33A^8 + 41A^9 -  $ \\  & $ 49A^{10} + 42A^{11} - 33A^{12} + 13A^{13} + 3A^{14} - 15A^{15} + 15A^{16} - 9A^{17} + 3A^{18}$ \\ \hline
 \hline\hline
\end{tabular}
\end{center}
\end{table}}


\subsection{Case Study 4: Circular graph representation}\label{CS4}
\begin{figure}
		\centering
		\includegraphics[width=3.5in]{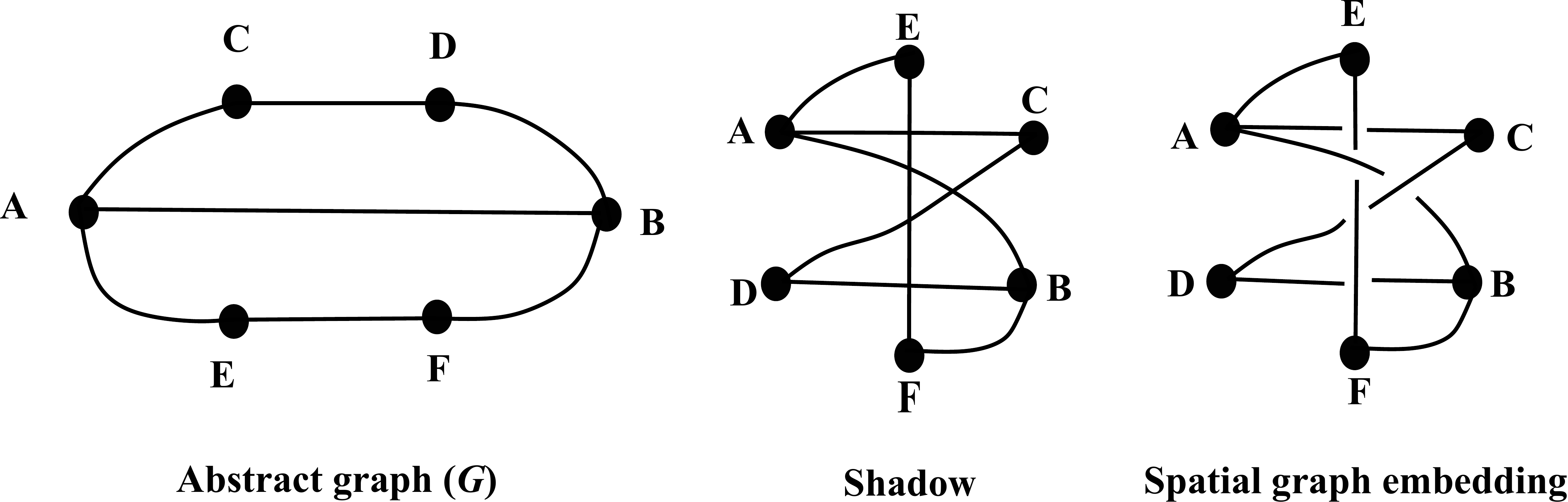}
		\caption{Implementation of the circular spatial graph representation technique in Sec.~\ref{CS4} to avoid unnecessary or extra twists between any two edges of a diagram.}
		\label{CaseStudy4}
\end{figure}

While filtering out isotopic spatial graph diagrams in the previous case studies, we observed that in a few occasions where system topologies have many crossings, two edges in that diagram twist around each other multiple times. Although a higher crossing number is satisfied, unnecessary intertwining between edges can often be reduced by Reidemeister moves to a smaller crossing number, so essentially, no unique spatial topology is attained. In some cases, such intertwining is practically not observed or desirable between pipes or ducts in most complex systems (e.g., aero-engine externals, hydraulic systems). Some syntactic constraints need to be imposed to prevent more than a simple crossing between any two edges. This requires a representation that implicitly forbids twisting of two edges multiple times around each other. One way to get different spatial embeddings of an input abstract graph $G$  as shown in Fig.~\ref{CaseStudy4} is to: 1) Pick an ordering of the nodes and use that to arrange them along a circle on the plane, 2) Connect the nodes by straight lines corresponding to the edges of $G$.  This gives the “shadow”, and 3) Resolve the intersections lines of the shadow into over or under crossings. Figure~\ref{CaseStudy4} shows the shadow of graph $G$ based on a particular cyclic order of nodes and one spatial graph embedding. As there are five crossings, a total of $2^{5} = 32$ spatial embeddings are possible. The unique ones can be identified using the proposed design framework. 


%

\subsection{Case Study 5: Large-scale system --- spatial graph decomposition approach}\label{CS5}
\begin{figure}
	\centering
		\includegraphics[width=4.5in]{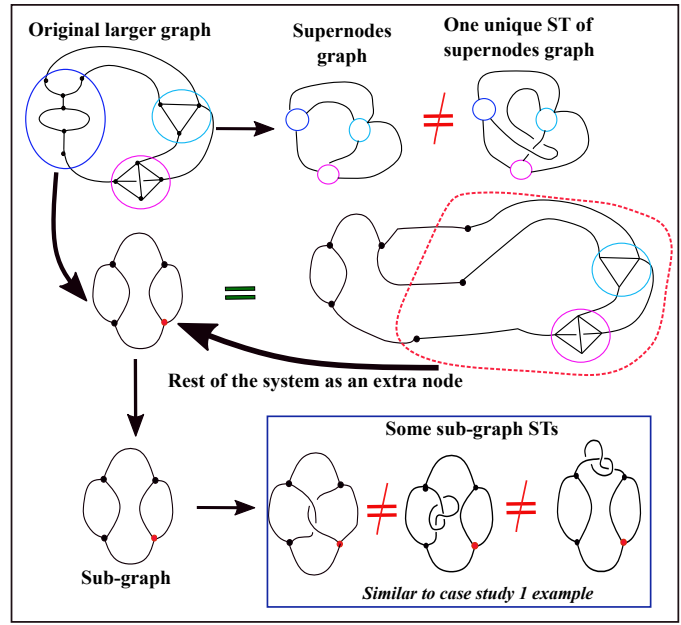}
		\caption{Demonstration of spatial graph diagram decomposition approach discussed in case study~\ref{CS5}.}
		\label{Casestudy5}
\end{figure}

From the observations made in the previous case studies, it is evident that enumerating spatial topologies for most real-world systems containing many components and approximately hundreds of crossings is intractable with manual processes and can become computationally expensive with automated methods such as the one presented in this article. In contrast, enumerating spatial topologies of each subsystem of components can be a simple and efficient process. A complex spatial graph can thus be converted to a set of sub-graphs, and the unique spatial topologies of these sub-graphs can then be enumerated separately. The sub-graphs can be decoupled and can be considered as super-nodes. This decouples the task into two subtasks: 1) Enumerate STs of the system graph with only the subsystems as super-nodes, and 2) enumerate unique STs within each sub-graph. This presents fewer design candidates about which to make decisions, which greatly reduces the overall computational expense. Figure~\ref{Casestudy5} shows a random complex spatial graph with 14 nodes, 20 edges and allowing at most 10 edge crossings. Approximately $1.134\times 10^{4}$ SGDs are attained for this entire system that fall under 434 unique Yamada polynomial categories. As this is a very large set, decomposition of the graph into sub-graphs (as super-nodes) is appropriate. First, a unique spatial topology of the super-nodes graph is found. Case study 1 in Sec.~\ref{CS1} is utilized as a sub-graph for demonstration purposes. Note that while enumerating STs for the spatial sub-graph, the rest of the system is condensed as an extra node in the sub-graph to preserve spatial connectivity information. Finally, using the proposed design framework, unique STs of the sub-graph can be plugged into the original system to attain system configurations. The scope of this article only deals with enumerating unique STs, so we plan to show how each of these unique topologies affect overall system performance in future work.\\ 

To explain this decomposition concept using a concrete engineering design example, suppose that the complex network represents the spatial topology of a hybrid-electric vehicle powertrain; one possible subsystem could be a fluid-thermal cooling circuit. Each distinct circuit topology can be geometrically optimized for fair comparison, revealing how the topological features contribute to the overall system efficiency, fuel economy, thermal loss management, and other figures of merit due to physical interactions between components, interconnections, and the environment. The best candidate ST can then be chosen according to the desired performance requirements, as was done in Refs.~\cite{Herber2019, Peddada2019a}, where the same procedure was followed but with the goal of ranking different system architectures (SAs). As part of future work on this topic, we plan to investigate alternative methods, such as deep learning and pattern recognition as mentioned in Refs.~\cite{Guo2018a, Davies2021, Song2022}, to efficiently explore new topologies of large-scale systems where exhaustive enumeration of all possible topologies may not be tractable.

\section{DISCUSSION}\label{Discussion}
 As described in Sec~\ref{RQ}, this article attempts to answer several fundamental research questions (not considered thoroughly in existing literature) on the design automation of spatial topological decisions and their implications on engineering system design. Based on the above case studies, several design insights were obtained. This section summarizes a list of important observations made from the five case studies as follows from an engineering design perspective and how those inferences help answer some research questions to a reasonable extent:
\begin{enumerate}
\item Firstly, the enumeration framework support representation of 3D systems as spatial graphs and allows an abstract representation of systems as grpahs with nodes and edges. Especially from case studies \ref{CS1} and \ref{CS2}, it can be observed that the number of unique spatial topologies attained for a given interconnect crossing complexity is much smaller than the combinatorially enumerated set of spatial graph diagrams, as most of them are isotopic to each other under the Reidemeister moves. Furthermore, implementation of the framework for enumeration of fuel cell system topologies in \ref{CS2} helps to identify non-obvious designs, thus providing value to creative design efforts for real-world engineering systems. This is a very promising result as navigation of 3D topological space efficiently is now possible and initial design filtering would help explore only useful candidate solutions. 
\item Another contribution of this work is embodied in Case study~\ref{CS3}, where STs are enumerated for components with different valencies, in contrast to existing work ~\cite{Oyamaguchi2015EnumerationOS, Kanenobu2012FiniteTI, Moriuchi2008EnumerationOA, Moriuchi2009ATO, Soma1996SpatialgraphIF} that is mostly limited to two or three equivalent vertices. This would be very useful in extending the framework to systems of various scales and requirements, including hollow objects and void regions. AI and ML techniques can be easily used through the data generated from such diverse design solutions.
\item  The circular graph representation method, presented in case study~\ref{CS4}, is a simple way to enumerate and realize SGDs and avoid edge intertwining, although Yamada polynomials should still be used for identifying unique STs.  Furthermore, specific syntactic constraints can be added to significantly  reduce the initial set of SGDs obtained for planarity checking and Yamada polynomial evaluation. For example, by adding constraints on total crossings allowed between two edges of a system, there is greater control on the type of spatial topologies finally obtained. This will be studied more in future work. Imposing syntactic constraints will be very useful to address the issue of interconnects having different physics spatial fields and optimal design trajectories for accessibility. Such constraints would be very useful to eliminate infeasible designs and either introduce or remove crossings between interconnects to satisfy operational, physics, or accessibility requirements. 
\item As seen in Case study~\ref{CS5}, for large-scaled systems, the best way so far to achieve different STs and search effectively is by graph decomposition. The spatial graph of the subsystem, which plays a critical role in performance impact, can be extracted to find its unique topologies. This avoids the need to enumerate thousands of diagrams of a complex network, compute their polynomials, and compare them. Moreover, sub-graph designs can be optimized for performance independently and then combined with the remaining system. In the decomposition-based approach, the isotopy equivalence check is first performed on sub-graphs and the then on the system graph containing the supernodes of each sub-graph. This case study is very helpful is addressing the question if it is possible to address subsystem level design challenges. For example, if certain portion of a larger system needs to be replaced or accessed for repair. Then graph decomposition would be very helpful to see accessibility scenarios for that subsystem with respect to the entire larger system and similarly module-based repair is possible and cost for overhaul becomes less expensive.
\item  Another impactful aspect of this framework is that for one system architecture, there can be a range of non-obvious spatial topologies from zero to many crossings. The spatial embeddings with fewer interconnect crossings are generally useful for practical engineering purposes where reduction of system geometric complexity is desired. Therefore, for existing, complex real-world system designs with 10s of crossings, using the proposed design framework, a simpler spatial topology may be found for that network with a much lower crossing number, but still keeping the same system connectivity. However, sometimes there might exist an implicit design requirement where having multiple interconnect crossings supports a system function. For example, applications involving wire harnesses, twisted cable bundles, etc., may benefit from additional crossings to support tighter space constraints or provide mechanical support. Hence, this framework can be utilized to satisfy different design requirements.
\end{enumerate}  

Several research questions such as a new design representation for 3D systems, scalability, time complexity, operational and maintenance considerations, simple geometry, crossing numbers, design automation for graph generation, etc. have been addressed in this article and the scope for further research on this topic is now expanded. In summary, advantages of using a spatial graph representation are: 1) its simplicity, while capturing necessary system elements and topological features, 2) ease of visualization, 3) flexibility to add new geometric features like size and shape, 4) the ability to detect distinct topologies using polynomial invariants, 5) scalable or even decomposable into a set of smaller graphs, 6) supports automated 3D model generation, and 7) accommodates features such as node locations, edge diameters, edge trajectory shape functions, port locations, crossing information, and other elements that can be parameterized for performing continuous numerical optimization. For example, items 6 and 7 can be very useful for using different 3D models as initial start points for multi-physics component placement and routing optimization. The features of the spatial graph design representation address several fundamental research questions however further investigation on this topic is needed to address the limitations of this framework and find out solutions to complex design issues. \\

\section{CONCLUSION}\label{Conclusion}
The design representation presented in this article greatly enhances the study of unique 3D engineering system spatial topologies in a systematic manner and is supported by rigorous mathematical foundations in spatial graph theory. Topologies of complex engineering systems, designed for particular applications, are conventionally created manually. But for more effective performance and design process efficiency, systematic identification, enumeration, and classification of possible system topologies can aid thorough navigation through challenging 3D discrete design spaces. A framework for representing three-dimensional interconnected engineering systems using spatial graph embeddings is presented. Initially, all the combinatorial spatial graph descriptions up to some fixed topological complexity are enumerated for an input system architecture. A polynomial invariant, the Yamada polynomial, is then calculated for the set of all the spatial graphs attained from the combinatorial permutations. The Yamada polynomial helps identify duplicate spatial graph topologies from the exhaustive set and a smaller set of unique spatial embeddings (equivalent topological classes) is obtained. This smaller set of spatial graphs can be used for generating three dimensional geometric system models. Five case studies have been demonstrated using the proposed enumeration strategy, including implementation of an industry application, an automotive fuel cell system (AFCS). The results show that this method is efficient, scalable, applicable to all general 3D interconnected system networks, allows comprehensive exploration of the design space, and greatly aids in the design and development of unprecedented system topologies.\\ 

Future work includes adding more geometric features to these spatial graph embeddings, such as representing nodes with geometric shapes and ports. Investigation of braid-based representations of interconnect networks is also anticipated. As designed systems become larger, evaluating Yamada polynomials for many SGDs is very time-consuming. This can be overcome by implementing a mix of Reidemeister moves to eliminate isotopic diagrams quickly to produce a smaller set of diagrams that require Yamada calculations. Other application aspects include utilizing the unique spatial topologies obtained here as starting points for physics-based component placement and routing optimization of 3D systems. Furthermore, research areas that can benefit from SGD representations are 3D pipe routing, topological 3D path planning for robotic operations, aerial drone navigation, generation of new automotive cooling system configurations, 3D integrated circuit interconnect technology, and many others. In this article, the general concept of 3D spatial topology enumeration using spatial graphs is discussed. We hope that this initial work serves as a strong foundation to bridging the gap between engineering design and mathematical low-dimensional topology. There are many interesting aspects which are yet to be explored and can have a great impact when applied to practical engineering design problems.

\section*{ACKNOWLEDGEMENTS}
 This material is based upon work supported by the National Science Foundation Engineering Research Center (NSF ERC) for Power Optimization of Electro-Thermal Systems (\href{http://poets-erc.org/}{\mbox{POETS}}) with cooperative agreement EEC-1449548. The opinions, findings, and conclusions or recommendations expressed are those of the author(s) and do not necessarily reflect the views of the National Science Foundation. Dunfield was partially supported by NSF grant DMS-1811156 and the Simons Foundation. The authors also acknowledge industry partners from Ford Motor Company who provided the representative data for the automotive fuel cell system (AFCS). The authors also thank Kyle Miller from the University of California, Berkeley for his helpful discussions on topics related to spatial graphs and Yamada polynomials. 

\bibliographystyle{asmems4}


\footnotesize{
\bibliography{refs}}

\begin{thebibliography}{100}

\bibitem{Peddada2021}
Peddada, S. R.~T., Dunfield, N.~M., Zeidner, L.~E., James, K.~A., and Allison,
  J.~T., 2021.
\newblock ``{Systematic Enumeration and Identification of Unique Spatial
  Topologies of 3D Systems Using Spatial Graph Representations}''.
\newblock In International Design Engineering Technical Conferences and
  Computers and Information in Engineering Conference, Vol.~Volume 3A: 47th
  Design Automation Conference (DAC).
\newblock V03AT03A042. \doi{10.1115/DETC2021-66900}.

\bibitem{Goode1957}
Goode, H.~H., and Machol, R.~E., 1957.
\newblock {\em System engineering: an introduction to the design of large-scale
  systems}.
\newblock McGraw-Hill New York.
\newblock \url{https://nla.gov.au/nla.cat-vn727826}.

\bibitem{Sydenham2003}
Sydenham, P.~H., 2003.
\newblock {\em Systems approach to engineering design}.
\newblock Artech House Boston, MA.
\newblock \url{https://nla.gov.au/nla.cat-vn2607407}.

\bibitem{Field2006}
Field, B.~W., 2006.
\newblock {\em Introduction to engineering design}.
\newblock University of Melbourne, Dept. of Mechanical Engineering, Clayton,
  Vic.
\newblock \url{https://nla.gov.au/nla.cat-vn3585408}.

\bibitem{Kim1991}
Kim, J.~J., and Gossard, D.~C., 1991.
\newblock ``Reasoning on the location of components for assembly packaging''.
\newblock {\em J. Mech. Des, \textbf{ 113}}(4), Dec., pp.~402--407.
\newblock \doi{10.1115/1.2912796}.

\bibitem{Ashrafiuon1990}
Ashrafiuon, H., and Mani, N.~K., 1990.
\newblock ``Analysis and optimal design of spatial mechanical systems''.
\newblock {\em J. Mech. Des, \textbf{ 112}}(2), June, pp.~200--207.
\newblock \doi{10.1115/1.2912593}.

\bibitem{Zhang2011}
Zhang, W., Xia, L., Zhu, J., and Zhang, Q., 2011.
\newblock ``Some recent advances in the integrated layout design of
  multicomponent systems''.
\newblock {\em J. Mech. Des, \textbf{ 133}}(10), Oct.
\newblock \doi{10.1115/1.4005083}.

\bibitem{Sergiy2018}
Yakovlev, S., and Kartashov, O., 2018.
\newblock ``{System Analysis and Classification of Spatial Configurations}''.
\newblock In 2018 IEEE First International Conference on System Analysis
  Intelligent Computing (SAIC), pp.~1--4.
\newblock \doi{10.1109/SAIC.2018.8516760}.

\bibitem{Liu2019}
Liu, F., Zhang, Y., Zheng, C., Qin, X., and Eynard, B., 2019.
\newblock ``Survey of configuration design approaches: A focus on design of
  complex industrial manufacturing systems''.
\newblock {\em Procedia CIRP, \textbf{ 81}}, pp.~340--345.
\newblock
  \doi{https://www.sciencedirect.com/science/article/pii/S2212827119303646}.

\bibitem{Blouin2004}
Blouin, V., Miao, Y., Zhou, X., and Fadel, G., 2004.
\newblock ``An assessment of configuration design methodologies''.
\newblock Multidisciplinary Analysis Optimization Conferences. American
  Institute of Aeronautics and Astronautics, Aug.
\newblock \doi{10.2514/6.2004-4430}.

\bibitem{Snavely1993}
Snavely, G.~L., and Papalambros, P.~Y., 1993.
\newblock Abstraction as a configuration design methodology, Sept.
\newblock \doi{10.1115/DETC1993-0317}.

\bibitem{Jiang2021}
Jiang, X., Hu, J., Peng, H., and Chen, Z., 2021.
\newblock ``A design methodology for hybrid electric vehicle powertrain
  configurations with planetary gear sets''.
\newblock {\em J. Mech. Des, \textbf{ 143}}(8), Feb.
\newblock \doi{10.1115/1.4049341}.

\bibitem{Schmidt1998}
Schmidt, L.~C., and Cagan, J., 1998.
\newblock ``Optimal configuration design: An integrated approach using
  grammars''.
\newblock {\em J. Mech. Des, \textbf{ 120}}(1), Mar., pp.~2--9.
\newblock \doi{10.1115/1.2826672}.

\bibitem{Deng2022}
Deng, T., Gan, Z., Xu, H., Wu, C., Zhang, Y., Liu, F., Ding, Z., and Chen, W.,
  2022.
\newblock ``Configuration design and screening of multi-mode
  double-planetary-gears hybrid powertrains''.
\newblock {\em J. Mech. Des, \textbf{ 144}}(7), Feb.
\newblock \doi{10.1115/1.4053525}.

\bibitem{Kott1992}
Kott, A., Agin, G., and Fawcett, D., 1992.
\newblock ``Configuration tree solver: A technology for automated design and
  configuration''.
\newblock {\em J. Mech. Des, \textbf{ 114}}(1), Mar., pp.~187--195.
\newblock \doi{10.1115/1.2916915}.

\bibitem{Campbell2000}
Campbell, M.~I., Cagan, J., and Kotovsky, K., 2000.
\newblock ``Agent-based synthesis of electromechanical design configurations''.
\newblock {\em J. Mech. Des, \textbf{ 122}}(1), Jan., pp.~61--69.
\newblock \doi{10.1115/1.533546}.

\bibitem{Grignon2004}
Grignon, P.~M., and Fadel, G.~M., 2004.
\newblock ``A ga based configuration design optimization method''.
\newblock {\em J. Mech. Des, \textbf{ 126}}(1), Mar., pp.~6--15.
\newblock \doi{10.1115/1.1637656}.

\bibitem{Sigurdarson2022}
Sigurdarson, N.~S., Eifler, T., Ebro, M., and Papalambros, P.~Y., 2022.
\newblock ``A novel approach to configuration redesign: Using multiobjective
  monotonicity analysis to alter the pareto set''.
\newblock {\em J. Mech. Des, \textbf{ 144}}(6), Feb.
\newblock \doi{10.1115/1.4053524}.

\bibitem{Bayrak2016}
Bayrak, A.~E., Ren, Y., and Papalambros, P.~Y., 2016.
\newblock ``Topology generation for hybrid electric vehicle architecture
  design''.
\newblock {\em J. Mech. Des, \textbf{ 138}}(8), June.
\newblock \doi{10.1115/1.4033656}.

\bibitem{Ai2005}
Ai, X., and Anderson, S., 2005.
\newblock An electro-mechanical infinitely variable transmission for hybrid
  electric vehicles.
\newblock \doi{10.4271/2005-01-0281}.

\bibitem{Ramdan2016}
Ramdan, M.~I., and Stelson, K.~A., 2016.
\newblock ``Optimal design of a power-split hybrid hydraulic bus''.
\newblock {\em Proceedings of the Institution of Mechanical Engineers, Part D:
  Journal of Automobile Engineering, \textbf{ 230}}(12), Jan., pp.~1699--1718.
\newblock \doi{10.1177/0954407015621817}.

\bibitem{Herber2019}
Herber, D.~R., and Allison, J.~T., 2019.
\newblock ``A problem class with combined architecture, plant, and control
  design applied to vehicle suspensions''.
\newblock {\em J. Mech. Des, \textbf{ 141}}(10), May.
\newblock \doi{10.1115/1.4043312}.

\bibitem{Shim1998}
Shim, P.~Y., and Manoochehri, S., 1998.
\newblock ``Optimal configuration design of structures using the binary
  enumeration technique''.
\newblock {\em Finite Elements in Analysis and Design, \textbf{ 31}}(1),
  pp.~15--32.
\newblock
  \url{https://www.sciencedirect.com/science/article/pii/S0168874X98000456}.

\bibitem{Gut2004}
Gut, J. A.~W., and Pinto, J.~M., 2004.
\newblock ``Optimal configuration design for plate heat exchangers''.
\newblock {\em International Journal of Heat and Mass Transfer, \textbf{
  47}}(22), pp.~4833--4848.
\newblock
  \url{https://www.sciencedirect.com/science/article/pii/S0017931004002170}.

\bibitem{Martins2009}
Martins, D., and Simoni, R., 2009.
\newblock ``Enumeration of planar metamorphic robots configurations''.
\newblock In 2009 ASME/IFToMM International Conference on Reconfigurable
  Mechanisms and Robots, pp.~580--588.
\newblock \url{https://ieeexplore.ieee.org/document/5173887}.

\bibitem{Peddada2019a}
Peddada, S. R.~T., Herber, D.~R., Pangborn, H.~C., Alleyne, A.~G., and Allison,
  J.~T., 2019.
\newblock ``{Optimal Flow Control and Single Split Architecture Exploration for
  Fluid-Based Thermal Management}''.
\newblock {\em Journal of Mechanical Design, \textbf{ 141}}(8), 04.
\newblock \doi{10.1115/1.4043203}.

\bibitem{Peddada2018Con}
Peddada, S. R.~T., Herber, D.~R., Pangborn, H.~C., Alleyne, A.~G., and Allison,
  J.~T., 2018.
\newblock ``Optimal flow control and single split architecture exploration for
  fluid-based thermal management''.
\newblock In ASME 2018 International Design Engineering Technical Conferences
  and Computers and Information in Engineering Conference, no.~V02AT03A005.
\newblock \doi{10.1115/DETC2018-86148}.

\bibitem{Peddada2020_JMD_2Stage}
Peddada, S. R.~T., James, K.~A., and Allison, J.~T., 2020.
\newblock ``{A Novel Two-Stage Design Framework for Two-Dimensional Spatial
  Packing of Interconnected Components}''.
\newblock {\em Journal of Mechanical Design, \textbf{ 143}}(3), 11.
\newblock 031706. \doi{10.1115/1.4048817}.

\bibitem{Bhattacharyya2022_AIAA}
Bhattacharyya, A., Peddada, S. R.~T., Bello, W.~B., Zeidner, L.~E., Allison,
  J.~T., and James, K.~A.
\newblock ``Simultaneous 3d component packing and routing optimization using
  geometric projection''.
\newblock In AIAA SCITECH 2022 Forum.
\newblock \doi{10.2514/6.2022-2096}.

\bibitem{Jessee2020a}
Jessee, A., Peddada, S. R.~T., Lohan, D.~J., James, K.~A., and Allison, J.~T.,
  2020.
\newblock ``Simultaneous packing and routing optimization using geometric
  projection''.
\newblock {\em ASME Journal of Mechanical Design}, Apr 2020.
\newblock \doi{10.1115/1.4046809}.

\bibitem{Peddada2022a}
Peddada, S., Zeidner, L., Ilies, H.~T., James, K., and Allison, J.~T., 2022.
\newblock ``Toward holistic design of spatial packaging of interconnected
  systems with physical interactions (spi2)''.
\newblock {\em J. Mech. Des}, July, pp.~1--26.
\newblock \url{https://doi.org/10.1115/1.4055055}.

\bibitem{Bayrak2016b}
Bayrak, A.~E., Kang, N., and Papalambros, P.~Y.
\newblock ``{Decomposition-Based Design Optimization of Hybrid Electric
  Powertrain Architectures: Simultaneous Configuration and Sizing Design}''.
\newblock 071405.

\bibitem{Muenzer2017a}
Muenzer, C., and Shea, K., 2017.
\newblock ``{Simulation-Based Computational Design Synthesis Using Automated
  Generation of Simulation Models From Concept Model Graphs}''.
\newblock {\em Journal of Mechanical Design, \textbf{ 139}}(7), 05.
\newblock 071101.

\bibitem{Field1997}
Field, B.~W., 1997.
\newblock {\em Introduction to design processes}, 2nd~ed.
\newblock Department of Mechanical Engineering, Monash University Clayton, Vic.
\newblock \url{https://nla.gov.au/nla.cat-vn556118}.

\bibitem{Challender2000}
Challender, S., 2000.
\newblock ``{Systems Thinking, Systems Practice. By Peter B. Checkland.
  Published by John Wiley, Chichester, UK, 1981, 330 pp., (republished 1999 in
  paperback, with a 30-year retrospective). From a practitioner perspective}''.
\newblock {\em Systems Research and Behavioral Science, \textbf{ 17}}(S1),
  pp.~S78--S80.
\newblock \doi{10.1002/1099-1743(200011)17:1+<::AID-SRES384>3.0.CO;2-N}.

\bibitem{Wyatt2013a}
Wyatt, D.~F., Wynn, D.~C., and Clarkson, P.~J., 2013.
\newblock ``A scheme for numerical representation of graph structures in
  engineering design''.
\newblock {\em J. Mech. Des}(1), Nov.
\newblock \doi{10.1115/1.4025961}.

\bibitem{Schmidt1999}
Schmidt, L.~C., Shetty, H., and Chase, S.~C., 1999.
\newblock ``A graph grammar approach for structure synthesis of mechanisms''.
\newblock {\em J. Mech. Des, \textbf{ 122}}(4), July, pp.~371--376.
\newblock \doi{10.1115/1.1315299}.

\bibitem{Oraon2018}
Oraon, N., and Rao, M., 2018.
\newblock ``Stick diagram representation for nanomagnetic logic based
  combinational circuits''.
\newblock In 2018 IEEE 18th International Conference on Nanotechnology
  (IEEE-NANO), pp.~420--425.
\newblock \doi{10.1109/NANO.2018.8626369}.

\bibitem{Babai1983CanonicalLO}
Babai, L., and Luks, E.~M., 1983.
\newblock ``Canonical labeling of graphs''.
\newblock {\em Proceedings of the fifteenth annual ACM symposium on Theory of
  computing}.
\newblock \doi{10.1145/800061.808746}.

\bibitem{Rensink2004}
Rensink, A., 2004.
\newblock ``Canonical graph shapes''.
\newblock In Programming Languages and Systems, D.~Schmidt, ed., Springer
  Berlin Heidelberg, pp.~401--415.
\newblock \doi{10.1007/978-3-540-24725-8\_28}.

\bibitem{Ross1991}
Ross, C.~S., and Route, W.~D., 1991.
\newblock ``A method for selecting parallel-connected, planetary gear train
  arrangements for automotive automatic transmissions''.
\newblock {\em SAE Transactions, \textbf{ 100}}, pp.~1765--1774.
\newblock \doi{http://www.jstor.org/stable/44632155}.

\bibitem{Liao2017}
Liao, Y.~G., and Chen, M.-Y., 2017.
\newblock ``Analysis of multi-speed transmission and electrically continuous
  variable transmission using lever analogy method for speed ratio
  determination''.
\newblock {\em Advances in Mechanical Engineering, \textbf{ 9}}(8), Aug.,
  p.~1687814017712948.
\newblock \doi{10.1177/1687814017712948}.

\bibitem{AlGeddawy2015}
AlGeddawy, T., and ElMaraghy, H., 2015.
\newblock ``{Determining Granularity of Changeable Manufacturing Systems Using
  Changeable Design Structure Matrix and Cladistics}''.
\newblock {\em Journal of Mechanical Design, \textbf{ 137}}(4), 04.
\newblock 041702. \doi{10.1115/1.4029515}.

\bibitem{Pease1988}
Pease, G., and Henderson, J.~M., 1988.
\newblock ``Simulation of a hydraulic hybrid vehicle using bond graphs''.
\newblock {\em J. Mech., Trans., and Automation, \textbf{ 110}}(3), Sept.,
  pp.~365--369.
\newblock \doi{10.1115/1.3267472}.

\bibitem{Wu2008}
Wu, Z., Campbell, M.~I., and Fernández, B.~R., 2008.
\newblock ``Bond graph based automated modeling for computer-aided design of
  dynamic systems''.
\newblock {\em J. Mech. Des, \textbf{ 130}}(4), Mar.
\newblock \doi{10.1115/1.2885180}.

\bibitem{Beaman1988}
Beaman, J.~J., and Rosenberg, R.~C., 1988.
\newblock ``Constitutive and modulation structure in bond graph modeling''.
\newblock {\em J. Dyn. Sys., Meas., Control, \textbf{ 110}}(4), Dec.,
  pp.~395--402.
\newblock \doi{10.1115/1.3152702}.

\bibitem{Behbahani2013}
Behbahani, S., and de~Silva, C.~W., 2013.
\newblock ``Automated identification of a mechatronic system model using
  genetic programming and bond graphs''.
\newblock {\em J. Dyn. Sys., Meas., Control, \textbf{ 135}}(5), May.
\newblock \doi{10.1115/1.4024171}.

\bibitem{Bachrach1996}
Bachrach, B.~I., 1996.
\newblock ``Annotated bond graphs--a communication tool''.
\newblock {\em J. Dyn. Sys., Meas., Control, \textbf{ 118}}(4), Dec.,
  pp.~797--800.
\newblock \doi{10.1115/1.2802361}.

\bibitem{Xu2019}
Xu, X., Sun, H., Liu, Y., and Dong, P., 2019.
\newblock ``Automatic enumeration of feasible configuration for the dedicated
  hybrid transmission with multi-degree-of-freedom and multiplanetary gear
  set''.
\newblock {\em J. Mech. Des, \textbf{ 141}}(9), Apr.
\newblock \doi{10.1115/1.4042846}.

\bibitem{Barhoumi2017}
Barhoumi, T., and Kum, D., 2017.
\newblock ``Automatic enumeration of feasible kinematic diagrams for split
  hybrid configurations with a single planetary gear''.
\newblock {\em J. Mech. Des, \textbf{ 139}}(8), May.
\newblock \doi{10.1115/1.4036583}.

\bibitem{Lipkin1991}
Lipkin, H., and Pohl, E., 1991.
\newblock ``Enumeration of singular configurations for robotic manipulators''.
\newblock {\em J. Mech. Des, \textbf{ 113}}(3), Sept., pp.~272--279.
\newblock \doi{10.1115/1.2912779}.

\bibitem{Liu1993}
Liu, T.~S., and Chou, C.~C., 1993.
\newblock ``Type synthesis of vehicle planar suspension mechanism using graph
  theory''.
\newblock {\em J. Mech. Des, \textbf{ 115}}(3), Sept., pp.~652--657.
\newblock \doi{10.1115/1.2919240}.

\bibitem{Sharma2014_VLSI_OptimizationSurvey}
Sharma, N., and Kaur, M., 2014.
\newblock ``A survey of vlsi techniques for power optimization and estimation
  of optimization''.
\newblock \doi{https://ijetae.com/files/Volume4Issue9/IJETAE\_0914\_50.pdf}.

\bibitem{Devadas1995ASO}
Devadas, S., and Malik, S., 1995.
\newblock ``A survey of optimization techniques targeting low power vlsi
  circuits''.
\newblock {\em 32nd Design Automation Conference}, pp.~242--247.
\newblock \doi{10.1109/DAC.1995.250098}.

\bibitem{Agnesina2020VLSIPP}
Agnesina, A., Chang, K., and Lim, S.~K., 2020.
\newblock ``Vlsi placement parameter optimization using deep reinforcement
  learning''.
\newblock {\em 2020 IEEE/ACM International Conference On Computer Aided Design
  (ICCAD)}, pp.~1--9.
\newblock \url{https://ieeexplore.ieee.org/document/9256814}.

\bibitem{BakerAlawieh2019GenerativeLI}
Alawieh, M.~B., Lin, Y., Ye, W.-C., and Pan, D.~Z., 2019.
\newblock ``Generative learning in vlsi design for manufacturability: Current
  status and future directions''.
\newblock {\em Journal of Microelectronic Manufacturing}.
\newblock \doi{10.33079/jomm.19020401}.

\bibitem{Bunglowala2009OptimizationOH}
Bunglowala, A., Singhi, B.~M., and Verma, A.~K., 2009.
\newblock ``Optimization of hybrid and local search algorithms for standard
  cell placement in vlsi design''.
\newblock {\em 2009 International Conference on Advances in Recent Technologies
  in Communication and Computing}, pp.~826--828.
\newblock \doi{10.1109/ARTCom.2009.163}.

\bibitem{Jung2016OpenDesignFD}
Jung, J., Jiang, I. H.-R., Nam, G.-J., Kravets, V.~N., Behjat, L., and Li,
  Y.-L., 2016.
\newblock ``Opendesign flow database: The infrastructure for vlsi design and
  design automation research''.
\newblock {\em 2016 IEEE/ACM International Conference on Computer-Aided Design
  (ICCAD)}, pp.~1--6.
\newblock \doi{10.1145/2966986.2980074}.

\bibitem{Nath2015ANA}
Nath, S., Ghosh, S., and Sarkar, S.~K., 2015.
\newblock ``A novel approach to discrete particle swarm optimization for
  efficient routing in vlsi design''.
\newblock {\em 2015 4th International Conference on Reliability, Infocom
  Technologies and Optimization (ICRITO) (Trends and Future Directions)},
  pp.~1--4.
\newblock \doi{10.1109/ICRITO.2015.7359375}.

\bibitem{Geetha2017DesignMA}
Geetha, B.~T., Padmavathi, B., and Perumal, V., 2017.
\newblock ``Design methodologies and circuit optimization techniques for low
  power cmos vlsi design''.
\newblock {\em 2017 IEEE International Conference on Power, Control, Signals
  and Instrumentation Engineering (ICPCSI)}, pp.~1759--1763.
\newblock \doi{10.1109/ICRITO.2015.7359375}.

\bibitem{Kumar2020ReviewOV}
Kumar, S. B.~V., Rao, P.~V., Sharath, H.~A., Sachin, B.~M., Ravi, U., and
  Monica, B.~V., 2020.
\newblock ``Review on vlsi design using optimization and self-adaptive particle
  swarm optimization''.
\newblock {\em J. King Saud Univ. Comput. Inf. Sci., \textbf{ 32}},
  pp.~1095--1107.
\newblock \doi{10.1016/j.jksuci.2018.01.001}.

\bibitem{Dewan2020NPSeparateAN}
Dewan, M.~I., and Kim, D.~H., 2020.
\newblock ``Np-separate: A new vlsi design methodology for area, power, and
  performance optimization''.
\newblock {\em IEEE Transactions on Computer-Aided Design of Integrated
  Circuits and Systems, \textbf{ 39}}, pp.~5111--5122.
\newblock \doi{10.1109/TCAD.2020.2966551}.

\bibitem{Held2011CombinatorialOI}
Held, S., Korte, B., Rautenbach, D., and Vygen, J., 2011.
\newblock ``Combinatorial optimization in vlsi design''.
\newblock In Combinatorial Optimization - Methods and Applications.
\newblock \doi{https://doi.org/10.3233/978-1-60750-718-5-33}.

\bibitem{Kaeslin2014TopDownDV}
Kaeslin, H., 2014.
\newblock ``Top-down digital vlsi design: From architectures to gate-level
  circuits and fpgas''.
\newblock
  \url{https://www.sciencedirect.com/book/9780128007303/top-down-digital-vlsi-design}.

\bibitem{Peddada2021b}
Peddada, S. R.~T., Zeidner, L.~E., James, K.~A., and Allison, J.~T., 2021.
\newblock ``{An Introduction to 3D SPI2 (Spatial Packaging of Interconnected
  Systems With Physics Interactions) Design Problems: A Review of Related Work,
  Existing Gaps, Challenges, and Opportunities}''.
\newblock In International Design Engineering Technical Conferences and
  Computers and Information in Engineering Conference, Vol.~Volume 3B: 47th
  Design Automation Conference (DAC).
\newblock V03BT03A034. \doi{10.1115/DETC2021-72106}.

\bibitem{peddada2023automated}
Peddada, S. R.~T., 2023.
\newblock Automated interference-free layout generation methods for 2d
  interconnected engineering systems.
\newblock Tech. rep.
\newblock \url{https://www.ideals.illinois.edu/items/126471}.

\bibitem{Moguel2017}
Moguel, E., Conejero, J.~M., Sánchez-Figueroa, F., Hernández, J., Preciado,
  J.~C., and Rodríguez-Echeverría, R., 2017.
\newblock ``Towards the use of unmanned aerial systems for providing
  sustainable services in smart cities''.
\newblock {\em Sensors (Basel, Switzerland), \textbf{ 18}}(29280984), Dec.,
  p.~64.
\newblock \url{https://www.ncbi.nlm.nih.gov/pmc/articles/PMC5795605/}.

\bibitem{Yu2018}
Yu, J., Cao, Z., Cheng, M., and Pan, R., 2018.
\newblock ``Hydro-mechanical power split transmissions: Progress evolution and
  future trends''.
\newblock {\em Proceedings of the Institution of Mechanical Engineers, Part D:
  Journal of Automobile Engineering, \textbf{ 233}}(3), Jan., pp.~727--739.
\newblock \doi{10.1177/0954407017749734}.

\bibitem{Kyprianidis2017AnAT}
Kyprianidis, K., 2017.
\newblock ``An approach to multi-disciplinary aero engine conceptual design''.
\newblock In International Symposium on Air Breathing Engines, ISABE 2017,
  Manchester, United Kingdom.
\newblock
  \url{http://mdh.diva-portal.org/smash/get/diva2:1305927/FULLTEXT01.pdf}.

\bibitem{Park2010}
Park, S., and Jung, D., 2010.
\newblock ``Design of vehicle cooling system architecture for a heavy duty
  series-hybrid electric vehicle using numerical system simulations''.
\newblock {\em J. Eng. Gas Turbines Power, \textbf{ 132}}(9), June.
\newblock \doi{10.1115/1.4000587}.

\bibitem{HosteThistleWeeks1998}
Hoste, J., Thistlethwaite, M., and Weeks, J., 1998.
\newblock ``The first 1,701,936 knots''.
\newblock {\em Math. Intelligencer, \textbf{ 20}}(4), pp.~33--48.
\newblock \doi{10.1007/BF03025227}.

\bibitem{Burton2020}
Burton, B.~A., 2020.
\newblock ``{The Next 350 Million Knots}''.
\newblock In 36th International Symposium on Computational Geometry (SoCG
  2020), S.~Cabello and D.~Z. Chen, eds., Vol.~164 of {\em Leibniz
  International Proceedings in Informatics (LIPIcs)}, Schloss
  Dagstuhl--Leibniz-Zentrum f{\"u}r Informatik, pp.~25:1--25:17.
\newblock \doi{10.4230/LIPIcs.SoCG.2020.25}.

\bibitem{Oyamaguchi2015EnumerationOS}
Oyamaguchi, N., 2015.
\newblock ``Enumeration of spatial 2-bouquet graphs up to flat vertex
  isotopy''.
\newblock {\em Topology and its Applications, \textbf{ 196}}, pp.~805--814.
\newblock \doi{10.1016/j.topol.2015.05.049}.

\bibitem{Kanenobu2012FiniteTI}
Kanenobu, T., and Sugita, K., 2012.
\newblock ``Finite type invariants of order 3 for a spatial handcuff graph''.
\newblock {\em Topology and its Applications, \textbf{ 159}}, pp.~966--979.
\newblock \doi{10.1016/j.topol.2011.11.016}.

\bibitem{Moriuchi2008EnumerationOA}
Moriuchi, H., 2008.
\newblock ``Enumeration of algebraic tangles with applications to theta-curves
  and handcuff graphs''.
\newblock {\em Kyungpook Math. J., \textbf{ 48}}, pp.~337--357.
\newblock \doi{10.5666/KMJ.2008.48.3.337}.

\bibitem{Moriuchi2009ATO}
Moriuchi, H., 2009.
\newblock ``A table of {$\theta$}-curves and handcuff graphs with up to seven
  crossings''.
\newblock In {\em Noncommutativity and singularities}, Vol.~55 of {\em Adv.
  Stud. Pure Math.} Math. Soc. Japan, Tokyo, pp.~281--290.
\newblock \doi{10.2969/aspm/05510281}.

\bibitem{Soma1996SpatialgraphIF}
Soma, T., 1996.
\newblock ``Spatial-graph isotopy for trivalent graphs and minimally knotted
  embeddings''.
\newblock {\em Topology and its Applications, \textbf{ 73}}, pp.~23--41.
\newblock \doi{10.1016/0166-8641(96)00035-1}.

\bibitem{FominykhEtAl2016}
Fominykh, E., Garoufalidis, S., Goerner, M., Tarkaev, V., and Vesnin, A., 2016.
\newblock ``A census of tetrahedral hyperbolic manifolds''.
\newblock {\em Exp. Math., \textbf{ 25}}(4), pp.~466--481.
\newblock \doi{10.1080/10586458.2015.1114436}.

\bibitem{Davies2021}
Davies, A., Veličković, P., Buesing, L., Blackwell, S., Zheng, D., Tomašev,
  N., Tanburn, R., Battaglia, P., Blundell, C., Juhász, A., Lackenby, M.,
  Williamson, G., Hassabis, D., and Kohli, P., 2021.
\newblock ``Advancing mathematics by guiding human intuition with ai''.
\newblock {\em Nature, \textbf{ 600}}(7887), pp.~70--74.
\newblock \doi{10.1038/s41586-021-04086-x}.

\bibitem{Guo2018d}
Guo, T., Herber, D.~R., and Allison, J.~T., 2018.
\newblock ``Reducing evaluation cost for circuit synthesis using active
  learning''.
\newblock In ASME 2018 International Design Engineering Technical Conferences,
  no.~DETC2018-85654, p.~V02AT03A011.
\newblock \doi{10.1115/DETC2018-85654}.

\bibitem{Guo2019a}
Guo, T., Herber, D.~R., and Allison, J.~T., 2019.
\newblock ``Circuit synthesis using generative adversarial networks ({GANs})''.
\newblock In AIAA 2019 Science and Technology Forum and Exposition, no.~AIAA
  2019-2350.
\newblock \doi{10.2514/6.2019-2350}.

\bibitem{Guo2018PhD}
Guo, T., 2018.
\newblock ``On the use of machine learning with design optimization data for
  system topology design.''.
\newblock {Ph.D.} {Dissertation}, University of Illinois at Urbana-Champaign,
  Urbana, IL, USA.
\newblock \url{ http://hdl.handle.net/2142/101819}.

\bibitem{Parrott2023a}
Parrott, C., Peddada, S., Allison, J.~T., and James, K.
\newblock {\em Machine Learning Surrogates for Optimal 2D Spatial Packaging of
  Interconnected Systems with Physics Interactions (SPI2)}.
\newblock \doi{10.2514/6.2023-4375}.

\bibitem{Flapan2016RecentDI}
Flapan, E., Mattman, T.~W., Mellor, B., Naimi, R., and Nikkuni, R., 2017.
\newblock ``Recent developments in spatial graph theory''.
\newblock In {\em Knots, links, spatial graphs, and algebraic invariants},
  Vol.~689 of {\em Contemp. Math.} Amer. Math. Soc., Providence, RI,
  pp.~81--102.
\newblock \arXiv{1602.08122}.

\bibitem{Taylor2019AbstractlyPS}
Taylor, S.~A., 2019.
\newblock Abstractly planar spatial graphs.
\newblock \arXiv{1902.01719}.

\bibitem{Flapan2012SpatialGW}
Flapan, E., Mellor, B., and Naimi, R., 2012.
\newblock ``Spatial graphs with local knots''.
\newblock {\em Revista Matem{\'a}tica Complutense, \textbf{ 25}}, pp.~493--510.
\newblock \doi{https://doi.org/10.48550/arXiv.1010.0479}.

\bibitem{Liang1994}
Liang, C., and Mislow, K., 1994.
\newblock ``Classification of topologically chiral molecules''.
\newblock {\em J. Math. Chemistry, \textbf{ 15}}(1), pp.~245--260.
\newblock \doi{10.1007/BF01277563}.

\bibitem{Flapan2013}
Flapan, E., and Fletcher, W., 2013.
\newblock ``Intrinsic chirality of multipartite graphs''.
\newblock {\em J. Math. Chemistry, \textbf{ 51}}(7), pp.~1853--1863.
\newblock \doi{10.1007/s10910-013-0187-y}.

\bibitem{FLAPAN1995}
Flapan, E., 1995.
\newblock ``Rigidity of graph symmetries in the {$3$}-sphere''.
\newblock {\em J. Knot Theory Ramifications, \textbf{ 4}}(3), pp.~373--388.
\newblock \doi{10.1142/S0218216595000181}.

\bibitem{Mellor2018InvariantsOS}
Mellor, B., 2018.
\newblock Invariants of spatial graphs.
\newblock \arXiv{1812.08885}.

\bibitem{Fleming2006AnIT}
Fleming, T., and Mellor, B., 2007.
\newblock ``An introduction to virtual spatial graph theory''.
\newblock In Proceedings of the International Workshop on Knot Theory for
  Scientific Objects, A.~Kawauchi, ed., Vol.~1 of {\em OCAMI Studies}, Osaka
  Municipal Universities Press.
\newblock \url{https://arxiv.org/abs/math/0608739}.

\bibitem{Rapenne2000}
Rapenne, G., Crassous, J., Echegoyen, L.~E., Echegoyen, L., Flapan, E., and
  Diederich, F., 2000.
\newblock ``Regioselective one-step synthesis and topological chirality of
  trans-3, trans-3,trans-3 and e,e,e [60]fullerene-cyclotriveratrylene
  tris-adducts: Discussion on a topological meso-form''.
\newblock {\em HCA, \textbf{ 83}}(6), June, pp.~1209--1223.
\newblock \doi{10.1002/1522-2675(20000607)83:6<1209::AID-HLCA1209>3.0.CO;2-Y}.

\bibitem{Dale2017}
Dale, M. R.~T., 2017.
\newblock ``Spatial graphs''.
\newblock Cambridge University Press, Cambridge, pp.~191--221.
\newblock \doi{10.1017/9781316105450.010}.

\bibitem{Tunyasuvunakool2021}
Tunyasuvunakool, K., Adler, J., Wu, Z., Green, T., Zielinski, M., Žídek, A.,
  Bridgland, A., Cowie, A., Meyer, C., Laydon, A., Velankar, S., Kleywegt,
  G.~J., Bateman, A., Evans, R., Pritzel, A., Figurnov, M., Ronneberger, O.,
  Bates, R., Kohl, S. A.~A., Potapenko, A., Ballard, A.~J., Romera-Paredes, B.,
  Nikolov, S., Jain, R., Clancy, E., Reiman, D., Petersen, S., Senior, A.~W.,
  Kavukcuoglu, K., Birney, E., Kohli, P., Jumper, J., and Hassabis, D., 2021.
\newblock ``Highly accurate protein structure prediction for the human
  proteome''.
\newblock {\em Nature, \textbf{ 596}}(7873), pp.~590--596.
\newblock \doi{10.1038/s41586-021-03828-1}.

\bibitem{Song2022}
Song, B., Luo, X., Luo, X., Liu, Y., Niu, Z., and Zeng, X., 2022.
\newblock ``Learning spatial structures of proteins improves protein-protein
  interaction prediction''.
\newblock {\em Brief Bioinform}, Jan., p.~bbab558.
\newblock \doi{10.1093/bib/bbab558}.

\bibitem{Heal2018}
Heal, J.~W., Bartlett, G.~J., Wood, C.~W., Thomson, A.~R., and Woolfson, D.~N.,
  2018.
\newblock ``Applying graph theory to protein structures: an atlas of coiled
  coils''.
\newblock {\em Bioinformatics (Oxford, England), \textbf{ 34}}(29722888), Oct.,
  pp.~3316--3323.
\newblock \url{https://www.ncbi.nlm.nih.gov/pmc/articles/PMC6157074/}.

\bibitem{Huan2003MiningSM}
Huan, J., Wang, W., Bandyopadhyay, D., Snoeyink, J., Prins, J., and Tropsha,
  A., 2003.
\newblock ``Mining spatial motifs from protein structure graphs''.
\newblock
  \url{https://www.semanticscholar.org/paper/Mining-Spatial-Motifs-from-Protein-Structure-Graphs-Huan-Wang/7a54e79a37ff14a4533d5be57961f01fc6e092b2}.

\bibitem{Trace1983OnTR}
Trace, B., 1983.
\newblock ``On the {R}eidemeister moves of a classical knot''.
\newblock {\em Proc. Amer. Math. Soc., \textbf{ 89}}(4), pp.~722--724.
\newblock \doi{10.2307/2044613}.

\bibitem{Hass1998TheNO}
Hass, J., and Lagarias, J., 1998.
\newblock ``The number of {R}eidemeister moves needed for unknotting''.
\newblock {\em J. Amer. Math. Soc., \textbf{ 14}}, pp.~399--428.
\newblock \doi{10.48550/arXiv.math/9807012}.

\bibitem{Hayashi2005}
Hayashi, C., 2005.
\newblock ``The number of {R}eidemeister moves for splitting a link''.
\newblock {\em Mathematische Annalen, \textbf{ 332}}(2), pp.~239--252.
\newblock \doi{10.1007/s00208-004-0599-x}.

\bibitem{HassLagariasPippenger1999}
Hass, J., Lagarias, J.~C., and Pippenger, N., 1999.
\newblock ``The computational complexity of knot and link problems''.
\newblock {\em J. ACM, \textbf{ 46}}(2), pp.~185--211.
\newblock \doi{10.1145/301970.301971}.

\bibitem{lackenby2019efficient}
Lackenby, M., to appear.
\newblock ``The efficient certification of knottedness and {T}hurston norm''.
\newblock {\em Adv. Math.}
\newblock \arXiv{1604.00290}.

\bibitem{Ishii2011ONNO}
Ishii, A., 2011.
\newblock ``On normalizations of a regular isotopy invariant for spatial
  graphs''.
\newblock {\em International J. Math., \textbf{ 22}}, pp.~1545--1559.
\newblock \doi{https://doi.org/10.1142/S0129167X1100729X}.

\bibitem{Negami1987PolynomialIO}
Negami, S., 1987.
\newblock ``Polynomial invariants of graphs''.
\newblock {\em Trans. Amer. Math. Soc., \textbf{ 299}}, pp.~601--622.
\newblock \doi{10.2307/2000516}.

\bibitem{Cho2011TopologicalSG}
Cho, S., and Koda, Y., 2011.
\newblock ``Topological symmetry groups and mapping class groups for spatial
  graphs''.
\newblock {\em Michigan Math. J., \textbf{ 62}}, pp.~131--142.
\newblock \doi{10.1307/MMJ/1363958244}.

\bibitem{Flapan2017KnotsLS}
Flapan, E., Henrich, A., Kaestner, A., and Nelson, S., eds., 2017.
\newblock Knots, links, spatial graphs, and algebraic invariants, Vol.~689 of
  {\em Contemporary Mathematics}, American Mathematical Society, Providence,
  RI.
\newblock \doi{10.1090/conm/689}.

\bibitem{BarNatan1995OnTV}
Bar-Natan, D., 1995.
\newblock ``On the {V}assiliev knot invariants''.
\newblock {\em Topology, \textbf{ 34}}, pp.~423--472.
\newblock \doi{10.1016/0040-9383(95)93237-2}.

\bibitem{Kauffman1988NewII}
Kauffman, L., 1988.
\newblock ``New invariants in the theory of knots''.
\newblock {\em Amer. Math. Monthly, \textbf{ 95}}, pp.~195--242.
\newblock \doi{10.1080/00029890.1988.11971990}.

\bibitem{Thompson1992API}
Thompson, A., 1992.
\newblock ``A polynomial invariant of graphs in 3-manifolds''.
\newblock {\em Topology, \textbf{ 31}}, pp.~657--665.
\newblock \doi{10.1016/0040-9383(92)90056-N}.

\bibitem{AlexanderTopologicalIO}
Alexander, J.~W., 1928.
\newblock ``Topological invariants of knots and links''.
\newblock {\em Trans. Amer. Math. Soc., \textbf{ 30}}(2), pp.~275--306.
\newblock \doi{10.2307/1989123}.

\bibitem{Murasugi1987JonesPA}
Murasugi, K., 1987.
\newblock ``Jones polynomials and classical conjectures in knot theory''.
\newblock {\em Topology, \textbf{ 26}}, pp.~187--194.
\newblock \doi{10.1142/9789812798329\_0026}.

\bibitem{Kauffman1989InvariantsOG}
Kauffman, L., 1989.
\newblock ``Invariants of graphs in three-space''.
\newblock {\em Trans. Amer. Math. Soc., \textbf{ 311}}, pp.~697--710.
\newblock \doi{10.1090/S0002-9947-1989-0946218-0}.

\bibitem{Dobrynin2003OnTY}
Dobrynin, A., and Vesnin, A., 2003.
\newblock ``On the yoshinaga polynomial of spatial graphs''.
\newblock {\em Kobe J. Math., \textbf{ 20}}, pp.~31--37.
\newblock \url{http://www.math.kobe-u.ac.jp/jmsj/kjm/}.

\bibitem{Yokota1996TopologicalIO}
Yokota, Y., 1996.
\newblock ``Topological invariants of graphs in 3-space''.
\newblock {\em Topology, \textbf{ 35}}, pp.~77--87.
\newblock \doi{10.1016/0040-9383(95)00002-X}.

\bibitem{Kong2015ColoringsDA}
Mellor, B., Kong, T., Lewald, A., and Pigrish, V., 2016.
\newblock ``Colorings, determinants and {A}lexander polynomials for spatial
  graphs''.
\newblock {\em J. Knot Theory Ramifications, \textbf{ 25}}(4), pp.~1650019, 19.
\newblock \arXiv{1506.06083}.

\bibitem{Murakami1993TheYP}
Murakami, J., 1993.
\newblock ``The {Y}amada polynomial of spacial graphs and knit algebras''.
\newblock {\em Comm. Math. Physics, \textbf{ 155}}, pp.~511--522.
\newblock \doi{10.1007/BF02096726}.

\bibitem{Yamada1989}
{Y}amada, S., 1989.
\newblock ``An invariant of spatial graphs''.
\newblock {\em J. Graph Theory, \textbf{ 13}}(5), Nov, pp.~537--551.
\newblock \doi{10.1002/jgt.3190130503}.

\bibitem{Vesnin1996}
Vesnin, A., and Dobrynin, A., 1996.
\newblock ``The {Y}amada polynomial for graphs, embedded knot-wise into
  three-dimensional space''.
\newblock {\em Vychislitel’nye Sistemy, \textbf{ 155}}, Jan.

\bibitem{Li2018OnYP}
Li, M., Lei, F., Li, F., and Vesnin, A., 2018.
\newblock ``The {Y}amada polynomial of spatial graphs obtained by edge
  replacements''.
\newblock {\em J. Knot Theory Ramifications, \textbf{ 27}}(9), pp.~1842004, 19.
\newblock \arXiv{1801.09075}.

\bibitem{Deng2018TheGY}
Deng, Q., Jin, X., and Kauffman, L.~H., 2019.
\newblock ``The generalized {Y}amada polynomials of virtual spatial graphs''.
\newblock {\em Topology Appl., \textbf{ 256}}, pp.~136--158.
\newblock \arXiv{1806.06462}.

\bibitem{HopcroftTarjan}
Hopcroft, J., and Tarjan, R., 1974.
\newblock ``Efficient planarity testing''.
\newblock {\em J. Assoc. Comput. Mach., \textbf{ 21}}, pp.~549--568.
\newblock \doi{10.1145/321850.321852}.

\bibitem{Peddada2020}
Peddada, S. R.~T., Rodriguez, S.~B., James, K.~A., and Allison, J.~T., 2020.
\newblock ``{Automated Layout Generation Methods for 2D Spatial Packing}''.
\newblock In International Design Engineering Technical Conferences and
  Computers and Information in Engineering Conference, Vol.~Volume 11B: 46th
  Design Automation Conference (DAC).
\newblock V11BT11A013, \doi{10.1115/DETC2020-22627}.

\bibitem{Peddada2020b}
Peddada, S. R.~T., James, K.~A., and Allison, J.~T., 2020.
\newblock ``{A Novel Two-Stage Design Framework for 2D Spatial Packing of
  Interconnected Components}''.
\newblock In International Design Engineering Technical Conferences and
  Computers and Information in Engineering Conference, Vol.~Volume 11B: 46th
  Design Automation Conference (DAC).
\newblock V11BT11A032. \doi{10.1115/DETC2020-22695}.

\bibitem{Peddada2021PhD}
Peddada, S. R.~T., 2021.
\newblock ``A two-stage design framework for optimal spatial packaging of
  fluid-thermal systems.''.
\newblock {Ph.D.} {Dissertation}, University of Illinois at Urbana-Champaign,
  Urbana, IL, USA, May.
\newblock \url{http://hdl.handle.net/2142/110458}.

\bibitem{Guo2018a}
Guo, T., Herber, D.~R., and Allison, J.~T., 2018.
\newblock ``Reducing evaluation cost for circuit synthesis using active
  learning''.
\newblock In ASME 2018 International Design Engineering Technical Conferences
  and Computers and Information in Engineering Conference, no.~V02AT03A011.
\newblock \doi{10.1115/DETC2018-85654}.

\end{thebibliography}

\end{document}